\documentclass[pra,aps,nofootinbib,showpacs,showkeys]{revtex4}

\def\XXint#1#2#3{{\setbox0=\hbox{$#1{#2#3}{\int}$}
     \vcenter{\hbox{$#2#3$}}\kern-.5\wd0}}

\usepackage{graphicx}
\usepackage{bm}
\usepackage{amssymb}
\usepackage{amsmath}
\usepackage{euscript}

\begin{document}

\title{Quantum Theory of Flicker Noise in Metal Films}

\author{Kirill~A.~Kazakov}\email{kirill_kazakov@comtv.ru}

\affiliation{Department of Theoretical Physics,
Physics Faculty,\\
Moscow State University, $119899$, Moscow, Russian Federation}

\begin{abstract}
Flicker ($1/f^{\gamma}$) voltage noise spectrum is derived from
finite-temperature quantum electromagnetic fluctuations produced
by elementary charge carriers in external electric field. It is
suggested that deviations of the frequency exponent $\gamma$ from
unity, observed in thin metal films, can be attributed to quantum
backreaction of the conducting medium on the fluctuating field of
the charge carrier. This backreaction is described
phenomenologically in terms of the effective momentum space
dimensionality, $\EuScript{D}.$ Using the dimensional continuation
technique, it is shown that the combined action of the photon heat
bath and external field results in a $1/f^{\gamma}$-contribution
to the spectral density of the two-point correlation function of
electromagnetic field. The frequency exponent is found to be equal
to $1 + \delta,$ where $\delta = 3 - \EuScript{D}$ is a reduction
of the momentum space dimensionality. This result is applied to
the case of a biased conducting sample, and a general expression
for the voltage power spectrum is obtained which possesses all
characteristic properties of observed flicker noise spectra. The
range of validity of this expression covers well the whole
measured frequency band. Gauge independence of the power spectrum
is proved. It is shown that the obtained results naturally resolve
the problem of divergence of the total noise power. A detailed
comparison with the experimental data on flicker noise
measurements in metal films is given.
\end{abstract}

\maketitle

\section{Introduction}

As is well-known, power spectra of voltage fluctuations in all
conducting media exhibit a universal low-frequency behavior -- for
sufficiently small frequencies $f$ they scale as $1/f^{\gamma},$
with the frequency exponent $\gamma$ about unity. This asymptotic
behavior is a manifestation of the so-called flicker noise present
in any system containing free-like charged particle states
\cite{buck}. Although the value of $\gamma$ and an overall
proportionality factor in this power law depend on many factors
such as sample material and sample geometry, system temperature,
etc., there are some important characteristic properties possessed
by all flicker noise power spectra. Namely, it is well established
experimentally that the noise produced by a biased sample is
proportional to the voltage bias squared, and roughly inversely
proportional to its volume. The enigmatic property of the
$1/f$-spectrum is its unboundedness. Flicker noise is present in
the whole measured frequency band covering more than twelve
decades. Experiments show no flattening of the spectrum for
frequencies as low as $10^{-6}~{\rm Hz}.$ Although on the opposite
side of the spectrum the $1/f$-component is dominated by other
types of noise (it compares with thermal noise usually at $f\sim
1~{\rm Hz}$), it has been detected at frequencies as high as $10^6
{\rm Hz}.$

Despite numerous models suggested since its discovery eighty years
ago \cite{johnson}, the origin of flicker noise still remains an
open issue. There is a widespread opinion that this noise arises
from resistance fluctuations, which is quite natural taking into
account its dependence on the applied bias. It has been proposed
that the resistance fluctuations possessing the other properties
of flicker noise might result from temperature fluctuations
\cite{voss1,hsiang}, fluctuations of the charge carrier mobility
\cite{hooge1,hooge2,klein1,klein2,vandamme}, or of the number of
charge carriers
\cite{bell,mcwhorter,ziel1,ziel2,klein3,jones,yakimov}. All these
models, however, have restricted validity, because they involve
one or another assumption specific to the problem under
consideration. For instance, assuming that the resistance
fluctuations arise from temperature fluctuations, one has to
choose an appropriate spatial correlation of these fluctuations in
order to obtain the desired profile of the power spectrum. In
addition to that, some models involve artificial normalization of
the power spectrum, needed to come up with the observed noise
level.

Perhaps the main difficulty for theoretical explanation is the
unboundedness of flicker noise spectrum. There are many physical
mechanisms that generate noise whose power spectrum has the
$1/f$-profile in some frequency domain. But these domains are so
narrow in comparison with the whole measured band that the
corresponding mechanisms cannot be considered as the general
mechanisms of flicker noise generation. For instance, according to
Ref.~\cite{stephany}, defect motion in carbon conductors generates
noise with power spectrum close to the inverse frequency
dependence in the frequency range $10^3~{\rm Hz}$ to $10^4~{\rm
Hz},$ while outside this interval it switches to $1/f^2.$ At the
same time, the omnipresence of flicker noise and high universality
of its properties suggest that there must exist an equally
universal reason for its occurrence.

This source is naturally expected to have a quantum origin.
Although some of the models suggested so far do consider various
quantum effects as underlying mechanisms of flicker noise (such
as, for instance, trapping of charge carriers), it may well be
that its origin is to be sought at the most fundamental level.
Namely, it is plausible that the phenomenon of flicker noise has
its roots in the very quantum nature of interaction of elementary
charges with electromagnetic field. From this point of view, the
problem has been attacked by Handel \cite{handel1}, who suggested
that flicker noise is the result of low-energy photon emission
accompanying any scattering process, and is related to the
infrared divergence of the cross-section considered as a function
of the energy loss. Later, the argument was modified and the
so-called coherent quantum $1/f$ effect described \cite{handel2},
which is connected with the infrared properties of the dressed
electron propagator. Although Handel's theory has been severely
criticized in many respects \cite{tremblay}, it has found support
in independent investigations of Refs.~\cite{vliet,ziel}.

An essentially different quantum approach to the problem was
proposed recently in Refs.~\cite{kazakov1,kazakov2}. In this
approach, flicker noise is treated as originating from quantum
fluctuations of individual electric fields of charge carriers. As
was shown in detail in Ref.~\cite{kazakov2}, spectral density of
the two-point correlation function of the Coulomb field produced
by a charge carrier exhibits the above-mentioned characteristic
properties of flicker noise. Namely, the low-frequency asymptotic
of the spectral density is $1/f,$ the noise intensity induced by
external electric field is proportional to the field strength
squared, and inversely proportional to the spatial separation
between the field producing particle and the observation point. In
application to the case of a conducting sample, the low-frequency
asymptotic of the voltage fluctuation power spectrum is found to
be
\begin{eqnarray}\label{mainhomr}
C_U(f) = \frac{\eta}{2\pi}\frac{U^2_0}{f}\,, \quad \eta =
\frac{2\alpha^2}{3ec}g\mu T\,,
\end{eqnarray}
\noindent where $U_0$ is the voltage bias, $T$ the system
temperature, $\alpha$ the fine structure constant, $\mu$ the
charge carrier mobility, $c$ the speed of light in vacuum, and $g$
a geometrical factor which is roughly inversely proportional to
the sample size. The range of validity of this result turns out to
be notably wide: The term ``low-frequency asymptotic'' means that
for a given $T,$ Eq.~(\ref{mainhomr}) is valid for frequencies
satisfying $f\ll 10^{11}T\,{\rm Hz},$ with $T$ expressed in
$^{\circ}{\rm K}.$ Covering well the whole band where flicker
noise has ever been observed, this condition in particular sets no
low-frequency cutoff, implying that the $1/f$-spectrum extends
down to zero. The fact that the found asymptotic does not require
a low-frequency cutoff is the consequence of its oddness with
respect to frequency. As discussed in Ref.~\cite{kazakov2},
appearance of such contributions to the spectral density is
related to the inhomogeneity in time of fluctuations produced by
individual charge carriers, and provides a natural resolution to
the problem of divergence of the total noise power.

It was demonstrated in Ref.~\cite{kazakov2} that
Eq.~(\ref{mainhomr}) is in agreement with the experimental results
of $1/f$-noise measurements in metals. Naturally, in this
comparison only genuine $1/f$ noise data were used, i.e., the data
that fit the law $1/f^{\gamma}$ in which $\gamma = 1,$ within
experimental error. Experiments show, however, that generally
power spectra follow the one-over-f law only in sufficiently thick
samples, while in thin samples (films, whiskers) large deviations
of $\gamma$ (up to $\gamma = 1.5$) are often observed. The purpose
of this paper is to show that these deviations can be described
within the developed theory by taking into account backreaction of
the conducting medium on the fluctuating electric field produced
by a charged carrier. Staying within the one-particle picture of
flicker noise generation, developed in
Refs.~\cite{kazakov1,kazakov2}, this backreaction can be described
as an effective reduction of the momentum space dimensionality. It
turns out that this reduction can be naturally realized using the
well-known techniques of dimensional continuation \cite{wilson}.
It will be shown that the power spectrum of electromagnetic
fluctuations, continued in this way to $\gamma > 1,$ adequately
describes the observed properties of flicker noise spectra in
metal films.

The paper is organized as follows. The effect of the backreaction
on photon propagation in a conducting film is discussed in
Sec.~\ref{exchangeint}. In Sec.~\ref{heatbath}, the influence of
the heat bath on quantum propagation of electromagnetic and
charged field quanta is considered, and the contributions relevant
in the low-frequency regime are identified. The power spectral
functions of the Coulomb field and voltage fluctuations are
defined in Sec.~\ref{prelim2}, and written down in the form
convenient for explicit calculations. Power spectrum of
electromagnetic fluctuations in the presence of external electric
field is evaluated in Sec.~\ref{calcul}. The low-frequency
asymptotic of the voltage power spectrum is found to obey the
$1/f^{\gamma}$-law with $\gamma = 3 - \EuScript{D},$ where
$\EuScript{D}$ is the effective dimensionality of momentum space.
Application of the obtained result to solids and comparison with
experimental data is given in Sec.~\ref{applications}. Some
important auxiliary material used in the main text is collected in
two appendices. Appendix~A contains the proof of gauge
independence of the voltage correlation function. Derivation of
the dimensionally reduced photon propagator in the axial gauge is
given in Appendix~B.

Units in which $\hbar = c = 1$ are used throughout this paper
except the end of Sec.~\ref{calcul} and Sec.~\ref{applications}.
The spacetime metric $\eta_{\mu\nu}$ is defined mostly negative,
i.e., $\eta_{\mu\nu} = {\rm diag}\{+1,-1,-1,-1\}.$

\section{Preliminaries}

\subsection{Photon propagation in a conducting film}\label{exchangeint}

Consider a conducting film, i.e., a sample which is thin in one,
say, $3$-direction. The film will be assumed homogeneous in
$(x_1,x_2)$-plane, but otherwise arbitrary. Let the film be in a
constant homogeneous electric field parallel to the
$(x_1,x_2)$-plane, and denote $U$ the voltage measured between two
leads attached to the film at the distance $L$ which is much
larger than the film thickness $a.$ We are interested in quantum
properties of the electromagnetic field produced by a charge
carrier moving in the film at finite temperature $T.$
Specifically, finite-temperature correlations in the values of the
particle's Coulomb field in the presence of external electric
field will be investigated. As long as backreaction of the
conducting medium on the fluctuating field of the charged particle
is neglected, these correlations are a single-particle effect, in
the sense that to the leading order in the electromagnetic
coupling, only fields produced by one and the same particle
correlate. However, this is no longer the case upon account of the
backreaction. The point is that under usual conditions of flicker
noise measurements, the backreaction is to be considered as an
effect of zero order in the electromagnetic coupling. Indeed,
since the characteristic time of the charge density response to
the field fluctuation is much smaller (normally, by ten orders of
magnitude at least) than the time $1/f$ of the field measurement,
this response can be considered quasi-stationary, leading to one
and the same charge density redistribution independently of the
value of charge carried by elementary medium constituents. Since
these redistributions tend to compensate the field fluctuation,
the backreaction leads to damping of the electric field
fluctuation. Concerning their effect on quantum propagation of the
particle's field, significance of these redistributions depends on
the mechanism governing the response. From this standpoint,
response of the medium through the ordinary electric conduction is
inconsequential because this is a classical effect, in the sense
that it relates quantities averaged over many decoherent particles
-- the mean electric field and mean electric charge density.
Things differ, however, upon account of specifically quantum
mechanisms such as the exchange interaction between charge
carriers. Since this interaction is independent of the particle
charge, it changes the structure of electromagnetic correlations
already at the lowest order. This interaction leads to
correlations of changes in the charge density distribution,
induced by the fluctuating electric field of the given particle.
Being related to the symmetry properties of the charged particles
state, the exchange interaction leads to density correlations
which are coherent, and therefore so is the response of the medium
on the fluctuating field. It is important that the charge carriers
correlated this way react coherently on each single photon
propagating in the medium, thus changing quantum properties of
this propagation. In other words, the exchange charge density
correlations modify the form of the electromagnetic field
propagator.

For definiteness, in the rest of this section we consider metal
conductors. Since electronic component in metals is degenerate at
ordinary temperatures, for a rough estimate of the exchange effect
it will be sufficient to use the simplest model of ideal
degenerate Fermi-gas of neutral particles. It is known
\cite{landau} that the exchange correlations of density
fluctuations in this case are destroyed by the thermal effects at
distances
$$r_0 = \frac{\hbar p_F}{2\pi m T}\,,$$ where $p_F$ is the
Fermi momentum, and $m$ the particle mass (for distances $r>r_0,$
density correlations fall off exponentially as $e^{-r/r_0}$). For
electrons in a metal, $p_F$ can be estimated as $\hbar/d,$
$d\approx 10^{-8}\,{\rm cm}$ being the lattice spacing, and hence
$$r_0 \approx  10^{-6}\,{\rm cm}\,.$$ At shorter distances, the correlation function
of density fluctuations scales as $r^{-4},$ and at $r \approx d$
fluctuations become completely correlated. Thus, one can expect
that the propagation of photons polarized in the $3$-direction
will be partially suppressed by the exchange effects in films with
thickness $a \lesssim r_0,$ and damped almost completely in the
case $a \approx d.$ This applies to the real transversely
polarized photons as well as to the longitudinal ``photons''
describing Coulomb fields of charged particles. Let us turn to the
consequences of this damping regarding the form of the
electromagnetic field propagator,
$$D_{\mu\nu}(x-x') = i\langle \EuScript{T}
\hat{A}_{\mu}(x)\hat{A}_{\nu}(x')\rangle\,,$$ where $\EuScript{T}$
denotes the usual time ordering of field operators, and averaging
is over the given photon state (equilibrium distribution at
temperature $T$). The quantum suppression of the $3$-component of
the electric field means that the electromagnetic field operator
$\hat{A}_{\mu}$ is constrained by the condition
$$\hat{E}_3 = - \frac{1}{c}\frac{\partial \hat{A_3}}{\partial t} -
\frac{\partial \hat{A}_0}{\partial x_3} = 0\,.$$ At the same time,
it is well known that the form of the photon propagator is to a
certain extent arbitrary. This arbitrariness reflects the freedom
in choosing gauge conditions used to fix the gradient invariance.
On the other hand, it is proved in Appendix~A that the voltage
correlation function is gauge-independent, so the choice of the
gauge is at our disposal. In the present case, it is convenient to
choose the axial gauge
\begin{eqnarray}\label{axial}
A_3 = 0\,.
\end{eqnarray}\noindent
Then the above operator constraint simplifies to
\begin{eqnarray}\label{azero}
\frac{\partial \hat{A}_0}{\partial x_3} = 0\,.
\end{eqnarray}\noindent This
condition means that the normal mode decomposition of the scalar
potential does not contain wave vectors with nonzero
$3$-component. Thus, in this particular gauge, the suppression of
$E_3$ implies that the ``temporal'' photons do not propagate in
the $3$-direction. In other words, the momentum space propagator
of these photons becomes effectively two-dimensional. It is
important, furthermore, that the complete photon propagator turns
out to be diagonal in the axial gauge under the condition
(\ref{azero}). Namely, the zero-temperature photon propagator is
found in Appendix~B to have the following form
\begin{eqnarray}\label{zerotphotonpr}
D^0_{\mu\nu}(x) = \left(\eta_{\mu\nu} +
n_{\mu}n_{\nu}\right)\frac{1}{a}\int\frac{d^3 k}{(2\pi)^3}
e^{-ikx}\frac{4\pi}{k^2 + i0}\,,
\end{eqnarray}
\noindent where $n_{\mu}$ is the unit vector in the $3$-direction.
As a consequence, the finite-temperature propagator is also
diagonal (see the next section). In view of this fact, $D_{00}$
turns out to be the only component of the photon propagator that
determines the power spectrum of voltage fluctuations produced by
nonrelativistic charge carriers. This is because whenever
$D_{\mu\nu}$ is contracted with the electromagnetic current
$J^{\nu},$ contribution of the components with $\nu \ne 0$ is
suppressed by the factor $|\bm{q}|/mc,$ where $\bm{q}$ is the
charged particle momentum, so that $D_{\mu\nu}J^{\nu}\approx
D_{\mu 0}J^0 = (D_{00}J^0,0,0,0).$ In what follows, therefore, we
will be dealing only with this component of the photon propagator,
and drop the lower Lorentz indices, for brevity.

Our choice of units $\hbar = c = 1$ still leaves freedom in
choosing the unit of length. It proves to be convenient to fix it
by setting $a=1,$ which will be assumed until the end of
Sec.~\ref{calcul}.

\subsection{Finite-temperature contribution to the propagators}\label{heatbath}

Let us now consider the influence of the heat bath on the photon
and charged particle propagation. As long as the mean electric
potential produced by an elementary charged particle is
considered, the photon heat bath has no effect. This is because
the 4-vector of momentum transfer to a free massive particle, $p,$
is always spacelike, $p^2<0,$ so the distribution of real photons
appearing in the definition of the photon propagator is
irrelevant. The consequence of this is that the quantities built
from the mean field, such as the disconnected part of the
correlation function, are not affected by the photon heat bath.
Things change, however, when the connected part of the two-point
correlation function of electric potential is considered. It is
defined by the following symmetric expression
\begin{eqnarray}\label{corr}&&
C^{\rm con}_{00}(x;x') = \frac{1}{2}\langle {\rm
in}|\hat{A}_0(x)\hat{A}_0(x') + \hat{A}_0(x')\hat{A}_0(x) |{\rm
in}\rangle\,,
\end{eqnarray}
\noindent where $x$ and $x'$ are the spacetime coordinates of two
observation points, $\hat{A}_0$ is the scalar potential operator
in the Heisenberg picture, and $|{\rm in }\rangle$ denotes the
given {\it in} state of the system ``charged particle +
electromagnetic field.'' In the two-photon processes, photon
momenta are allowed to take on lightlike directions, and hence the
photon heat bath does contribute to the function $C^{\rm
con}_{00}(x;x').$

As is well-known, the ordinary Feynman rules of the S-matrix
theory are not generally applicable for the calculation of {\it
in-in} expectation values, and must be modified, e.g., according
to Schwinger and Keldysh \cite{schwinger,keldysh}. This
complication was overcome in \cite{kazakov2} by rewriting
Eq.~(\ref{corr}) in the form
\begin{eqnarray}\label{corr2}&&
C^{\rm con}_{00}(x;x') = {\rm Re}\langle {\rm out}|T
\{\hat{A}_0(x)\hat{A}_0(x')\}|{\rm in}\rangle\,,
\end{eqnarray}
\noindent which allows the use of the S-matrix rules. This
transformation uses equivalence of the one-particle {\it in} and
{\it out} states (and the Hermiticity of the electromagnetic field
operator). Thus, in order to calculate the {\it in-in} expectation
value (\ref{corr}) taking into account the heat bath effect, the
standard finite-temperature-field-theory techniques can be used
\cite{landsman}. Below we employ a version of the real time
formalism, developed in \cite{niemi}, which is especially
convenient in actual calculations since the momentum space
propagators in this formulation do not involve the step function.

The real time formulation involves doubling of all fields, which
will be specified by a two-valued lower index. According to the
diagrammatic rules derived in \cite{niemi}, the photon propagator
at finite temperature can be obtained in momentum space (of
arbitrary dimensionality) by replacing the quantity $4\pi/(k^2 +
i0)$ in Eq.~(\ref{zerotphotonpr}) by the following matrix
\begin{eqnarray}\label{phiphot}
\mathfrak{D}(k) = 4\pi\left(
\begin{array}{cc}
D_{11}(k)&D_{12}(k)\\
D_{21}(k)&D_{22}(k)
\end{array}\right)\,,
\end{eqnarray}
\noindent where
\begin{eqnarray}
D_{11}(k) &=& - D^*_{22}(k) = \frac{1}{k^2 + i0} - \frac{2\pi i
\delta(k^2)}{e^{\beta|k_0|} - 1}\,, \nonumber \\ D_{12}(k) &=&
D_{21}(k) = - \frac{2\pi
i\delta(k^2)e^{\beta|k_0|/2}}{e^{\beta|k_0|} - 1}\,,\quad \beta =
1/T\,.\nonumber
\end{eqnarray}
\noindent In applications to the problem of $1/f$-noise considered
below, the value of the product $\beta |k_0|$ turns out to be very
small. For instance, even for frequencies as large as $10^6 {\rm
Hz},$ and temperatures as small as $1^{\circ}{\rm K}, $ it does
not exceed $\hbar\cdot 10^6/k \approx 10^{-5}$ ($k$ is the
Boltzmann constant), so the denominators in the above expressions
can be replaced by $\beta |k_0|,$ implying that the second term in
$D_{11}$ dominates. On the other hand, the temperature effect on
the propagation of massive particles is much less prominent. For
instance, in the case of conduction electrons in a crystal (this
case will be used throughout as a standard example), the particle
energy is of the order $(\hbar/d)^2/m.$ Setting $m\approx
10^{-27}{\rm g},$ we find that $(\hbar/d)^2/mT \approx 10^{+5}/T,$
with $T$ expressed in $^{\circ}{\rm K}.$ Hence, the temperature
contribution can be completely neglected (for fermions as well as
for bosons), and the propagator taken in the simple diagonal form
\begin{eqnarray}\label{phiprop}
\mathfrak{D}^{\phi}(x) &=& \int\frac{d^4 k}{(2\pi)^4}
\mathfrak{D}^{\phi}(k)e^{-ikx}\,, \quad \mathfrak{D}^{\phi}(k) =
\left(
\begin{array}{cc}
D^{\phi}_{11}(k)& 0\\
0& D^{\phi}_{22}(k)
\end{array}\right)\,, \\
D^{\phi}_{11}(k) &=& - D^{\phi *}_{22}(k) = (m^2 - k^2 -
i0)^{-1}\,.\nonumber
\end{eqnarray}
\noindent This is for a scalar particle described by the action
\begin{eqnarray}\label{action}
S_{\phi} &=& {\displaystyle\int} d^4 x
\left\{(\partial_{\mu}\phi^* + i e A_{\mu}\phi^*)
(\partial^{\mu}\phi - ie A^{\mu}\phi) - m^2 \phi^*\phi\right\}\,,
\end{eqnarray} \noindent where $e$ is the particle charge.
Account of particle spin, though adds some extra algebra, does not
change the long-range properties of its field, so that following
Ref.~\cite{kazakov2} we work with the simplest case of zero-spin
particles. The matrix propagators are multiplied in the
interaction vertices, generated by the triple and higher order
terms in the Lagrangian, with an additional minus sign for the
product of 2-components, as in the Schwinger-Keldysh techniques.
The arguments of the Green functions are treated as 1-component
fields. Finally, external particle lines represent normalized
particle amplitudes or their conjugates, according to whether the
particle is incoming or outgoing, just like in the conventional
techniques.

As was mentioned above, the photon propagator is dominated by the
temperature contribution (as long as the range of momentum
integration contains lightlike directions, see discussion in
Sec.~\ref{calcul}), while in the massive particle propagator this
contribution is negligible. It is important, on the other hand,
that the heat bath affects significantly the real particle
propagation, i.e., external matter lines in the diagrams.
Bilinears of the particle amplitudes representing these lines are
expressed eventually via statistical distribution function (see
Sec.~\ref{calcul} for details). Thus, apart from explicit
$T$-dependence coming from the photon propagator, the correlation
function also depends on temperature implicitly through the
particle statistical distribution.

\subsection{Power spectral densities of potential and voltage fluctuations}
\label{prelim2}

Connected contribution to the power spectral density of electric
potential fluctuations is obtained by Fourier transforming
Eq.~(\ref{corr}) with respect to the difference of the time
instants $t,t'$:
\begin{eqnarray}\label{corrf}&&
C(\bm{x},\bm{x}',t',\omega) = \int\limits_{-\infty}^{+\infty}d\tau
C^{\rm con}_{00}(\bm{x},t'+\tau;\bm{x}',t')e^{-i\omega\tau}\,.
\end{eqnarray}
\noindent The upper and lower indices in the notation of the
correlation function are suppressed in the left hand side, for
brevity. We are interested ultimately in the power spectrum of
voltage fluctuations, $C_U,$ measured between two observation
points $\bm{x},\bm{x}'$ (the two leads attached to the film). The
connected contribution to the voltage correlation function is
given by
\begin{eqnarray}\label{corru}&&
C_U(\bm{x},\bm{x}',t,t') = \frac{1}{2}\langle{\rm
in}|\hat{U}(t)\hat{U}(t') + \hat{U}(t')\hat{U}(t)|{\rm
in}\rangle\,,
\end{eqnarray}
\noindent where $\hat{U}(t) = \hat{A}_0(\bm{x},t) -
\hat{A}_0(\bm{x}',t)$ is the operator of voltage between the two
points. This function is separately symmetric with respect to the
interchanges $\bm{x}\leftrightarrow \bm{x}',$ and
$t\leftrightarrow t',$ unlike the function $C^{\rm
con}_{00}(x;x')$ which is only symmetric under $x\leftrightarrow
x'.$ Substituting the definition of $\hat{U}(t)$ in
Eq.~(\ref{corru}), the former can be expressed via the latter as
\begin{eqnarray}
C_U(\bm{x},\bm{x}',t,t') &=& C^{\rm con}_{00}(\bm{x},t;\bm{x},t')
+ C^{\rm con}_{00}(\bm{x}',t;\bm{x}',t') \nonumber \\ &-&
\left[C^{\rm con}_{00}(\bm{x},t;\bm{x}',t') + C^{\rm
con}_{00}(\bm{x}',t;\bm{x},t')\right]\,.
\end{eqnarray}
\noindent Accordingly, the power spectrum of voltage fluctuations,
defined by
\begin{eqnarray}\label{corruf}&&
C_U(\bm{x},\bm{x}',t',\omega) =
\int\limits_{-\infty}^{+\infty}d\tau
C_U(\bm{x},t'+\tau,\bm{x}',t')e^{-i\omega\tau}\,,
\end{eqnarray}
\noindent is expressed through that of potential fluctuations as
\begin{eqnarray}\label{corrfup}
C_U(\bm{x},\bm{x}',t',\omega) &=& C(\bm{x},\bm{x},t',\omega) +
C(\bm{x}',\bm{x}',t',\omega) - \left[C(\bm{x},\bm{x}',t',\omega) +
C(\bm{x}',\bm{x},t',\omega)\right]\,.
\end{eqnarray}
\noindent Although $C_U(\bm{x},\bm{x}',t,t')$ is symmetric with
respect to the interchange $t\leftrightarrow t',$ it depends on
both time arguments separately, and therefore
$C_U(\bm{x},\bm{x}',t,\omega)$ does not have to be an even
function of $\omega.$

When calculating the power spectrum of potential fluctuations
according to Eqs.~(\ref{corr2}), (\ref{corrf}), it is convenient
to perform the Fourier transformation under the sign ``{\rm Re}''
in Eq.~(\ref{corr2}). For this purpose, we introduce Fourier
transform of the two-point Green function:
\begin{eqnarray}\label{fourierg}&&
G(\bm{x},\bm{x}',t',\omega) = \int\limits_{-\infty}^{+\infty}d\tau
G(\bm{x},t'+\tau;\bm{x}',t') e^{-i\omega\tau}\,, \quad G(x;x') =
\langle {\rm out}|T \{\hat{A}_0(x)\hat{A}_0(x')\}|{\rm
in}\rangle\,, \nonumber \\
\end{eqnarray}
\noindent with the help of which the power spectrum of potential
fluctuations can be written as
\begin{eqnarray}\label{corrg}
C(\bm{x},\bm{x}',t',\omega) &=& \frac{1}{2}{\rm Re
}\left\{G(\bm{x},\bm{x}',t',\omega) +
G(\bm{x},\bm{x}',t',-\omega)\right\} \nonumber\\ &+&
\frac{i}{2}{\rm Im }\left\{G(\bm{x},\bm{x}',t',\omega) -
G(\bm{x},\bm{x}',t',-\omega)\right\}\,.
\end{eqnarray}
\noindent We see that contributions to the function
$C(\bm{x},\bm{x}',t',\omega),$ and hence to the voltage power
spectrum, can be either real even, or imaginary odd functions of
frequency.

\section{Low-frequency asymptotic of the power spectrum}\label{calcul}

\begin{figure}
\includegraphics{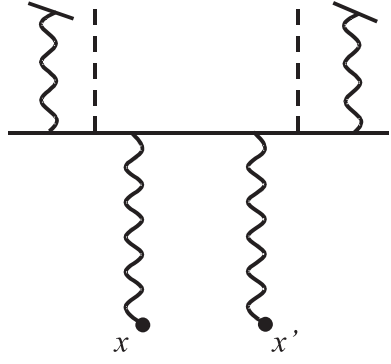}
\caption{Symbolic diagrammatic picture of the effect of particle
collisions and external electric field on the particle wave
function. Wavy lines denote photon propagators, broken line the
external electric field, solid lines charged particles.}
\label{fig1}
\end{figure}

\begin{figure}
\includegraphics{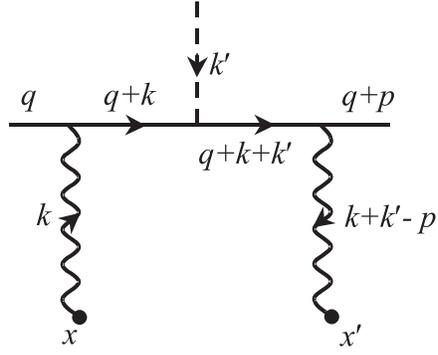}
\includegraphics{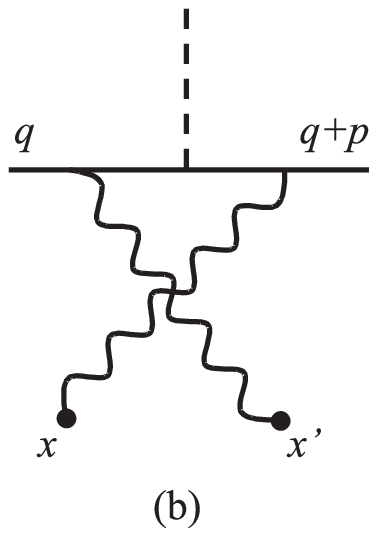}
\caption{Feynman diagrams describing the first order external
field correction to the particle propagator. $q$ and $p$ are the
particle 4-momentum and 4-momentum transfer, respectively.}
\label{fig2}
\end{figure}

In this section, spectral density of the voltage correlation
function will be evaluated in the low-frequency limit taking into
account the finite-temperature effect and the influence of
constant homogeneous external electric field. Let us first discuss
the role of the particle collisions in this calculation. It was
mentioned in Sec.~\ref{exchangeint} that, barring the exchange
interaction, correlations in the values of the electromagnetic
fields produced by the charge carriers is a one-particle effect
within the leading order in the electromagnetic coupling, in that
only fields produced by one and the same particle correlate. At
the same time, in the presence of the external electric field the
direct particle interactions cannot be neglected completely.
Indeed, although this field leads to a relatively small
corrections to the charge carrier wave function and its propagator
in all practically relevant cases, the effect of constant
homogeneous field on the free-like particles cannot be treated
perturbatively. The role of the particle collisions is to prevent
the charge carrier from gaining too much momentum from the field,
thus cutting down its effect. This is pictured schematically in
Fig.~\ref{fig1} where the particle collisions are symbolized  by a
virtual photon interchange between the particles. The external
field and the collisions affect both the particle wave function
and its propagator. As to the former, account of these two factors
is accomplished by replacing the particle momentum probability
distribution by the statistical distribution function, obtained as
a solution of the kinetic equation in the presence of external
electric field (this point will be discussed in more detail later
in this section). Regarding the particle propagator, however, the
issue is not that simple, because in order to extract the
low-frequency asymptotic of the power spectrum one needs an
explicit expression for the propagator, incorporating both
effects, which is unknown. This difficulty can be overcome by the
following trick. As was mentioned above, the external field gives
rise only to a relatively small correction to the particle
propagator. It is, therefore, sufficient to determine this
correction in the linear approximation, i.e., to the first order
with respect to the electromagnetic coupling. We now make an
assumption (confirmed by the result of the calculation) that the
low-frequency asymptotic of the power spectrum is independent of
the charge carrier mass. Then this asymptotic can be found
formally in the large mass limit. For a given frequency $\omega,$
one can always take the mass large enough so as to justify
perturbative treatment of the external field effect. To the first
order in the electromagnetic coupling, corrections to the particle
propagator due to external electric field and particle collisions
are well-separated from each other because they are superimposed
linearly. On the other hand, since the characteristic time of
particle collisions is very small, their contribution to the
particle propagator is inconsequential in calculating the
low-frequency asymptotic of the power spectrum. Thus, the lowest
order contribution to the correlation function is represented by
diagrams with a single insertion of the vertex describing
interaction of the charged particle with external electric field.
These diagrams are shown in Fig.~\ref{fig2}. Distinguishing the
contributions of the diagrams \ref{fig2}(a) and \ref{fig2}(b) by
the corresponding Latin subscript, we have
\begin{eqnarray}\label{diagen}&&
G_a(x,x') = -e^3\iiint d^4 z d^4 z' d^4 z''A_0(z'')\nonumber\\&&
\times\left\{\mathfrak{D}(x-z)\left[\psi(z)
\stackrel{\leftrightarrow}{\partial_{0}}
\mathfrak{D}^{\phi}(z-z'')\stackrel{\leftrightarrow}{\partial_{0}^{\,\prime\prime}}
\mathfrak{D}^{\phi}(z''-z')\stackrel{\leftrightarrow}{\partial_{0}^{\,\prime}}
\psi^*(z')\right]\mathfrak{D}(z'-x')\right\}_{11}\,, \\&&
G_b(x,x') = G_a(x',x)\,,\nonumber
\end{eqnarray}
\noindent where
\begin{eqnarray}
\varphi\stackrel{\leftrightarrow}{\partial_{0}}\psi &=&
\varphi\partial_{0}\psi - \psi\partial_{0}\varphi\,, \quad
\partial_{0}^{\,\prime} = \frac{\partial}{\partial z'_0}\,, \nonumber
\end{eqnarray}
\noindent $A_0(z'') = -(\bm{E}\bm{z}'')$ is the external field
potential,\footnote{We do not include an arbitrary constant in
this expression because it is clear in advance that it cannot
affect the final result. It is proved in Appendix A that the
correlation function is actually invariant under the most general
gauge variations of the electromagnetic potential.} and $\psi$ the
given particle state. Next, we go over to momentum space with the
help of Eqs.~(\ref{phiphot}), (\ref{phiprop}), and introduce the
spectral function for $G(x,x')$ according to Eq.~(\ref{fourierg}).
Assuming also that the charged particle is nonrelativistic, and
taking into account that $D^{\phi}_{12} = 0,$ we write the matrix
product longhand\footnote{The relation
$G_b(\bm{x},\bm{x}',t',\omega) = G^*_a(\bm{x},\bm{x}',t',-\omega)$
is conveniently proved using the following sequence of
substitutions in Eqs.~(\ref{diagenk1e}) -- (\ref{diagenk3e}):
$\bm{k}\to \bm{k} + \bm{p} - \bm{k}'\,,$ $\bm{q} \to \bm{q} -
\bm{p}\,,$ and then $\bm{p} \to - \bm{p}\,.$ The extra factor
$(-1)$ coming from the complex conjugation of the imaginary unit
in $A_0(\bm{k}')$ is compensated by that from the integration by
parts with respect to $\bm{k}'.$}
\begin{eqnarray}\label{diagenk1e}
G_a(\bm{x},\bm{x}',t',\omega) = &&\hspace{-0,3cm} (4\pi e)^2\iiint
\frac{d^3 \bm{q}}{(2\pi)^3} \frac{d^{\EuScript{D}}
\bm{p}}{(2\pi)^{\EuScript{D}}}\frac{d^{\EuScript{D}}
\bm{k}'}{(2\pi)^{\EuScript{D}}}\psi(\bm{q})\psi^*(\bm{q} + \bm{p})
\nonumber\\&& \times e^{ip^0(t' - t_0) - i(\bm{p} -
\bm{k}')\bm{x}'}A_0(\bm{k}') J_a(p,q,k',\bm{x}-\bm{x}',\omega)\,,
\\ G_b(\bm{x},\bm{x}',t',\omega) = &&\hspace{-0,3cm}
G^*_a(\bm{x},\bm{x}',t',-\omega)\,, \quad p^0 =
\frac{(\bm{q}+\bm{p})^2}{2m} - \frac{\bm{q}^2}{2m}\,,
\end{eqnarray}
\noindent where
\begin{eqnarray}\label{diagenk3e}
J_a(p,q,k',\bm{x}-\bm{x}',\omega) = && \hspace{-0,3cm}-
\left.ie(2m)^2\int \frac{d^{\EuScript{D}}
\bm{k}}{(2\pi)^{\EuScript{D}}}
e^{i\bm{k}(\bm{x}'-\bm{x})}\right.\\&& \left.
\times\left\{D_{11}(k)D^{\phi}_{11}(q+k)D^{\phi}_{11}(q+k+k')D_{11}(k+k'-p)
\right.\right.\nonumber\\&& \left.\left. \hspace{0,5cm} -
D_{12}(k)D^{\phi}_{22}(q+k)D^{\phi}_{22}(q+k+k')D_{21}(k+k'-p)\right\}
\right|_{k_0=\omega}, \nonumber\\ A_0(\bm{k}') = &&\hspace{-0,3cm}
- i(2\pi)^{\EuScript{D}} \left(\bm{E}\frac{\partial}{\partial
\bm{k}'}\right)\delta^{({\EuScript{D}})}(\bm{k}')\,, \quad k' =
(0,k_1',k_2',0)\,, \quad {\EuScript{D}} = 2. \nonumber
\end{eqnarray}
\noindent Here $q_{\mu}$ is the charged particle 4-momentum, and
$\psi(\bm{q})$ its momentum wave function at some time instant
$t_0,$ normalized by
\begin{eqnarray}\label{norm}
\int\frac{d^3 \bm{q}}{(2\pi)^3}|\psi(\bm{q})|^2 = 1\,.
\end{eqnarray}\noindent
That the integrations over momenta $\bm{k},\bm{p},\bm{k}'$ in the
expressions (\ref{diagenk1e}), (\ref{diagenk3e}) turned out to be
two-dimensional follows from the expression for the photon
propagator, derived in the preceding section, which in the current
notation can be written as
$$\mathfrak{D}(x) =
\int\frac{d^{\EuScript{D}+1}
k}{(2\pi)^{\EuScript{D}+1}}e^{-ikx}\mathfrak{D}(k)\,, \quad
\EuScript{D}=2\,,$$ $\mathfrak{D}$ being given by
Eq.~(\ref{phiphot}). As we saw in Sec.~\ref{exchangeint}, this
reduction of momentum space dimensionality from $\EuScript{D}=3$
to $\EuScript{D}=2$ is the consequence of quantum damping of
temporal photon propagation in the $3$-direction. This suppression
is noticeable only in sufficiently thin films with $a \lesssim
r_0,$ being formally complete in the practically unattainable case
$a = d,$ and is negligible in thick samples with $a \gg r_0.$ To
interpolate between these two limiting cases, it is natural to
consider $\EuScript{D}$ in the above formulas as a
phenomenological parameter allowed to take on non-integral values
between $2$ and $3,$ and which therefore can be called ``effective
dimensionality'' of the film. The difference $3-\EuScript{D}
\equiv \delta$ thus will be an effective reduction of the film
dimensionality, caused by the suppression. To accomplish this
interpolation, we employ the well-known technique of dimensional
continuation, developed by Wilson (in condensed matter theory),
and by t'Hooft and Veltman (in quantum field theory)
\cite{wilson}.

Before evaluating the $\EuScript{D}$-integrals, let us note that
the integrand in Eq.~(\ref{diagenk3e}) can be considerably
simplified. First of all, the second term in the curly brackets
can be neglected. Indeed, in view of the factor $D_{12}(k)$ which
is proportional to $\delta(k^2),$ and the condition $k_0 =
\omega,$ the momentum $k$ contributes only a tiny value to the
argument of the factor $D_{21};$ for electrons in a crystal, for
instance, the ratio $|\bm{k}|/|\bm{p}| \sim (\hbar
\omega/c)/(\hbar/L) = \omega L/c \approx 10^{-10}\omega L,$ where
$\omega,L$ are expressed in the $CGS$ system of units ($L$ is the
sample length). Even for $L$ as large as $1\,{\rm cm},$ this ratio
is very small for all practically relevant frequencies. Taking
into account also that the momentum $\bm{k}'$ is set eventually
equal to zero, the factor $D_{21}(k+k'-p)$ can be written simply
as $D_{21}(p)\sim \delta(p^2).$ But the argument of this
delta-function is always nonzero, because momentum transfer $p$ to
the massive particle is spacelike. Furthermore, using explicit
expression for the photon propagators, their product in the first
term in the curly brackets reads
\begin{eqnarray}&&
D_{11}(k)D_{11}(k+k'-p) = \frac{1}{[k^2 + i0]}\frac{1}{[(k+k'-p)^2
+ i0]} - \frac{2\pi i \delta(k^2)}{e^{\beta|k_0|} -
1}\frac{1}{(k+k'-p)^2 + i0}\nonumber\\&& \hspace{4cm} -
\frac{1}{k^2 + i0}\frac{2\pi i
\delta((k+k'-p)^2)}{e^{\beta|k_0-p_0|} - 1} + \frac{2\pi i
\delta(k^2)}{e^{\beta|k_0|} - 1}\frac{2\pi i
\delta((k+k'-p)^2)}{e^{\beta|k_0-p_0|} - 1}\,.\nonumber
\end{eqnarray}
\noindent As before, the last term in this expression can be
omitted, while the first term is dominated by the second, as we
saw in Sec.~\ref{heatbath}. Furthermore, $k_0$ enters the
temperature exponent in the third term in the combination $(k_0 -
p_0),$ and therefore, this term does not contribute to the leading
term of the low-frequency asymptotic, because in practice
$|\omega|\ll p_0.$ Indeed, estimating the energy transfer $p_0$ as
$(\bm{p}\bm{q})/m \approx \hbar^2/mdL,$ one finds for our standard
example $|\omega/p_0|\sim 10^{-8}|\omega|L.$ Finally, in the
product of the second term with
$D^{\phi}_{11}(q+k)D^{\phi}_{11}(q+k+k'),$ the pole of the
function $D^{\phi}_{11}(q+k)$ does not contribute because of the
factor $\delta(k^2).$ This is again a consequence of the
requirement that the momentum transfer to the massive particle
on-shell be spacelike: conditions $k^2 = 0,$ $q^2 = m^2,$ and
$(q+k)^2 = m^2$ cannot be satisfied altogether. Hence, the scalar
particle propagator in this product can be written simply as
$D^{\phi}_{11}(q+k) = - 1/2(qk) \approx - 1/2m\omega.$ Setting
also $(e^{\beta|\omega|} - 1)^{-1} \approx T/|\omega|,$ one thus
finds
\begin{eqnarray}
G_a(\bm{x},\bm{x}',t',\omega) = &&\hspace{-0,3cm} -
i(2\pi)^{\EuScript{D}}(4\pi e)^2\iiint \frac{d^3 \bm{q}}{(2\pi)^3}
\frac{d^{\EuScript{D}}
\bm{p}}{(2\pi)^{\EuScript{D}}}\frac{d^{\EuScript{D}}
\bm{k}'}{(2\pi)^{\EuScript{D}}}\psi(\bm{q})\psi^*(\bm{q} + \bm{p})
\nonumber\\&& \times e^{ip^0(t' - t_0) - i(\bm{p} -
\bm{k}')\bm{x}'}\left(\bm{E}\frac{\partial}{\partial
\bm{k}'}\right)\delta^{(\EuScript{D})}(\bm{k}')
J_a(p,q,k',\bm{x}-\bm{x}',\omega)\,,\nonumber
\end{eqnarray}
\noindent where
\begin{eqnarray}&&
J_a(p,q,k',\bm{x}-\bm{x}',\omega) = - \frac{4\pi e m
T}{\omega|\omega|}\int \frac{d^{\EuScript{D}}
\bm{k}}{(2\pi)^{\EuScript{D}}}
e^{i\bm{k}(\bm{x}'-\bm{x})}\left.\frac{D^{\phi}_{11}(q+k+k')
\delta(k^2)}{(k+k'-p)^2 + i0}\right|_{k_0 = \omega}\,. \nonumber
\end{eqnarray}
\noindent The singular at $\omega = 0$ contribution to the
function $G_a(\bm{x},\bm{x}',t',\omega)$ comes from integration
over small $\bm{k}.$ The strength of this singularity is
determined by the poles of the propagators in the integrand, and
we have to decide which of them gives rise to the strongest
singularity after performing the $\bm{k}'$-integration. Consider
first the case when the $\bm{k}'$-derivative acts on the factors
$[(k+k'-p)^2 + i0]^{-1},$ $e^{- i(\bm{p} - \bm{k}')\bm{x}'}.$
Since these depend on the difference $(\bm{p}-\bm{k}'),$ changing
$\partial/\partial \bm{k}' \to -
\partial/\partial \bm{p},$ and then integrating by parts with respect
to $\bm{p}$ in Eq.~(\ref{diagenk1e2}), this derivative is rendered
to act on terms independent of $\bm{k}.$ Thus, of all the
$\bm{k}'$-dependent factors in the integrand only
$D^{\phi}_{11}(q+k+k')$ is to be differentiated in effect. This
brings the above expression for $G_a$ to the form
\begin{eqnarray}
G_a(\bm{x},\bm{x}',t',\omega) = &&\hspace{-0,3cm} i(4\pi e)^2\iint
\frac{d^3 \bm{q}}{(2\pi)^3} \frac{d^{\EuScript{D}}
\bm{p}}{(2\pi)^{\EuScript{D}}}\psi(\bm{q})\psi^*(\bm{q} + \bm{p})
\nonumber\\&& \times e^{ip^0(t' - t_0) -
i\bm{p}\bm{x}'}\left(\bm{E}\bm{q}\right)
J_a(p,q,\bm{x}-\bm{x}',\omega) \nonumber
\end{eqnarray}
\noindent with
\begin{eqnarray}
J_a(p,q,\bm{x}-\bm{x}',\omega) = - \frac{2\pi e
T}{m\omega^3|\omega|\bm{p}^2}\int \frac{d^{\EuScript{D}}
\bm{k}}{(2\pi)^{\EuScript{D}}} e^{i\bm{k}(\bm{x}'-\bm{x})}
\delta(\omega^2 - \bm{k}^2)\,. \nonumber
\end{eqnarray}
\noindent It is seen from Eq.~(\ref{corrfup}) that if a term in
the function $C(\bm{x},\bm{x}',t',\omega)$ is independent of one
of the arguments $\bm{x},$ $\bm{x}',$ then it does not contribute
to the voltage power spectrum. Hence, we expand the exponent in
the integrand of (\ref{diagenk3e2}), and retain only one term
leading in the low-frequency limit. Using the rules of
$\EuScript{D}$-integration \cite{wilson}, one finds
\begin{eqnarray}\label{diagenk3e2}
J_a(p,q,\bm{x}-\bm{x}',\omega) &=& \frac{\pi e
T}{m\omega^3|\omega|\bm{p}^2}\int \frac{d^{\EuScript{D}}
\bm{k}}{(2\pi)^{\EuScript{D}}} [\bm{k}(\bm{x}'-\bm{x})]^2
\delta(\omega^2 - \bm{k}^2) \nonumber\\ &=& \frac{\pi e
T(\bm{x}'-\bm{x})^2}{m\omega^3|\omega|\bm{p}^2}\int
\frac{d^{\EuScript{D}} \bm{k}}{(2\pi)^{\EuScript{D}}}
\frac{\bm{k}^2}{\EuScript{D}} \delta(\omega^2 - \bm{k}^2)
\nonumber\\ &=& \frac{\pi e
T(\bm{x}'-\bm{x})^2}{m\omega^3|\omega|\bm{p}^2}
\frac{|\omega|^{\EuScript{D}}
S_{\EuScript{D}}}{2\EuScript{D}(2\pi)^{\EuScript{D}}}\,, \nonumber
\end{eqnarray}
\noindent where $$S_{\EuScript{D}} =
\frac{2\pi^{\EuScript{D}/2}}{\Gamma(\EuScript{D}/2)}$$ is the area
of unit hypersphere in $\EuScript{D}$ dimensional space
($\Gamma(x)$ is the Euler function). Thus,
\begin{eqnarray}
G_a(\bm{x},\bm{x}',t',\omega) =
\frac{i2\pi(4\pi)^{2-\EuScript{D}/2}}{\EuScript{D}\Gamma(\EuScript{D}/2)}\frac{e^3
TL^2}{2m\omega|\omega|^{\delta}}\iint \frac{d^3 \bm{q}}{(2\pi)^3}
\frac{d^{\EuScript{D}}
\bm{p}}{(2\pi)^{\EuScript{D}}}\psi(\bm{q})\psi^*(\bm{q} + \bm{p})
e^{ip^0(t' - t_0) -
i\bm{p}\bm{x}'}\frac{(\bm{E}\bm{q})}{\bm{p}^2}\,, \nonumber
\end{eqnarray}
\noindent where $L = |\bm{x}'-\bm{x}|.$ The total contribution to
the spectral density of the two-point Green function is
$$G(\bm{x},\bm{x}',t',\omega) = G_a(\bm{x},\bm{x}',t',\omega) +
G^*_a(\bm{x},\bm{x}',t',-\omega)\,.$$ Substituting this into
Eq.~(\ref{corrg}) one sees that since
$G_a(\bm{x},\bm{x}',t',\omega)$ is odd with respect to frequency,
it is the imaginary part of the Green function that contributes to
the correlation function $C(\bm{x},\bm{x}',t',\omega)$ which
therefore takes the form :
\begin{eqnarray}\label{diagenk1e2}
\hspace{-0,5cm}C(\bm{x},\bm{x}',t',\omega) =
\frac{i2\pi(4\pi)^{2-\EuScript{D}/2}}{\EuScript{D}\Gamma(\EuScript{D}/2)}\frac{e^3
TL^2}{m\omega|\omega|^{\delta}}{\rm Re}\iint \frac{d^3
\bm{q}}{(2\pi)^3} \frac{d^{\EuScript{D}}
\bm{p}}{(2\pi)^{\EuScript{D}}}\psi(\bm{q})\psi^*(\bm{q} + \bm{p})
e^{ip^0(t' - t_0) -
i\bm{p}\bm{x}'}\frac{(\bm{E}\bm{q})}{\bm{p}^2}\,. \nonumber
\hspace{-1cm}\\
\end{eqnarray}
\noindent In a many-particle system, this result is to be
expressed through the one-particle density matrix, which is
accomplished by replacing $\psi^*(\bm{q}')\psi(\bm{q}) \to
\varrho_0(\bm{q}',\bm{q}),$ where $\varrho_0$ is the momentum
space density matrix at the time instant $t_0.$ Recall that $t_0$
is the instant at which the particle state $\psi(\bm{q})$ is
prepared. It can be identified, for instance, as the moment the
charge carrier enters the sample, or escapes from a surface trap,
etc. The factor $e^{ip_0(t'-t_0)}$ in the integrand realizes
evolution of the density matrix from the instant $t_0$ to $t'.$
Since $p_0 = (\bm{q} + \bm{p})^2/2m - \bm{q}^2/2m,$ the product
$e^{ip_0(t'-t_0)}\varrho_0(\bm{q}+\bm{p},\bm{q})$ describes a
particle evolving freely on the interval $(t_0,t').$ On the other
hand, as was already mentioned in the beginning of this section,
in order to justify perturbative treatment of the external field
effect on the real particle states, it is necessary to take into
account particle collisions. For this purpose, it is sufficient to
consider these collisions as instantaneous. Then the interval
$(t_0,t')$ is divided into a sequence of short time intervals of
duration $\tau_f$ (the particle mean free time), on each of which
the density matrix evolves freely, and changes abruptly at the
collision instants. Going through this sequence, the density
matrix tends to the stationary statistical distribution function,
$\varrho(\bm{q}',\bm{q}),$ which is independent of the initial
particle state. What is important here is the sign of the
difference $(t'-t_0).$ Recall that $t'$ is a fixed time instant to
count off the time interval $\tau$ with respect to which the
correlation function is Fourier-transformed, and that each
particle has its own $t_0.$ This means that for a given $\omega,$
the system is observed during the time interval $(t' - \Delta t,
t' + \Delta t),$ where $\Delta t\sim 1/\omega,$ and $t_0$s are
distributed uniformly over this interval. The density matrix
evolves forward (backward) in time, if $t'>t_0$ ($t'<t_0$). But
time reversal involves inversion of particle momentum, and
therefore, the reciprocal contributions to the function
$G(\bm{x},\bm{x}',t',\omega)$ have opposite signs. To be more
specific, let $(t'-t_0)>0.$ Then the exponent $e^{ip_0(t'-t_0)}$
realizes forward evolution of the density matrix, so that the
integral in Eq.~(\ref{diagenk1e2}) takes eventually the form
\begin{eqnarray}\label{contrf}
\iint \frac{d^3 \bm{q}}{(2\pi)^3}\frac{d^{\EuScript{D}}
\bm{p}}{(2\pi)^{\EuScript{D}}} \varrho(\bm{q}+\bm{p},\bm{q})
\frac{(\bm{E}\bm{q})}{\bm{p}^2} e^{- i\bm{p}\bm{x}'}\,.
\end{eqnarray}
\noindent On the other hand, if $(t'-t_0)<0,$ then the density
matrix evolves backward. In momentum space, the initial state of
the reversed motion is represented by the amplitude
$\tilde{\psi}(\bm{q}) = \psi^*(-\bm{q}).$ Taking complex conjugate
of the integral in Eq.~(\ref{diagenk1e2}) (which does not change
the value of $G$ in view of the sign ``Re''), and changing the
integration variables $\bm{q}\to - \bm{q},$ $\bm{p}\to - \bm{p}$
gives in this case
$$ - \iint \frac{d^3 \bm{q}}{(2\pi)^3}\frac{d^{\EuScript{D}}
\bm{p}}{(2\pi)^{\EuScript{D}}}
\tilde{\psi}(\bm{q})\tilde{\psi}^*(\bm{q}+\bm{p})
\frac{(\bm{E}\bm{q})}{\bm{p}^2} e^{ip_0(t_0-t') -
i\bm{p}\bm{x}'}\,.$$ Replacing
$\tilde{\psi}^*(\bm{q}+\bm{p})\tilde{\psi}(\bm{q}) \to
\tilde{\varrho}(\bm{q}+\bm{p},\bm{q}),$ where $\tilde{\varrho}$
plays the role of momentum density matrix at the moment $t',$ the
exponent $e^{ip_0(t_0-t')}$ governs forward evolution of this
state on the interval $(t',t_0),$ so that the above expression
takes the form
$$ - \iint \frac{d^3 \bm{q}}{(2\pi)^3}\frac{d^{\EuScript{D}}
\bm{p}}{(2\pi)^{\EuScript{D}}} \varrho(\bm{q}+\bm{p},\bm{q})
\frac{(\bm{E}\bm{q})}{\bm{p}^2} e^{- i\bm{p}\bm{x}'}\,.$$ The
density matrix here is the same as in (\ref{contrf}), because the
statistical distribution is independent of the initial state. We
see that reciprocal contributions to the function
$G(\bm{x},\bm{x}',t',\omega)$ cancel each other when summed over
all particles in the system. Thus, we arrive at the important
conclusion that the total noise intensity is independent of the
number of particles, and remains at the level of individual
contribution. As was shown in Ref.~\cite{kazakov2}, this
conclusion is also true of the disconnected part of the
correlation function, though by virtue of quite different reasons.

It is customary to further express the function
$\varrho(\bm{q}',\bm{q})$ via the real mixed distribution
function, $n(\bm{r},\bm{q}),$ according to
\begin{eqnarray}
\varrho(\bm{q}+\bm{p},\bm{q}) = \int d^3\bm{r}e^{i(\bm{p}\bm{r})}
n\left(\bm{r},\bm{q} + \frac{\bm{p}}{2}\right)\,. \nonumber
\end{eqnarray}
\noindent Probability distributions for the particle position in a
sample or its momentum can be obtained by integrating
$n(\bm{r},\bm{p})$ over all $\bm{p}$ or the sample volume,
respectively. Using this in the expression (\ref{contrf}), and
substituting the latter into Eq.~(\ref{diagenk1e2}) yields
\begin{eqnarray}
C(\bm{x},\bm{x}',t',\omega) =
\frac{i2\pi(4\pi)^{2-\EuScript{D}/2}}{\EuScript{D}\Gamma(\EuScript{D}/2)}\frac{e^3
TL^2}{m\omega|\omega|^{\delta}}{\rm Re}\iiint \frac{d^3
\bm{q}}{(2\pi)^3}
\frac{d^{\EuScript{D}}\bm{p}}{(2\pi)^{\EuScript{D}}}d^3\bm{r}
\frac{(\bm{E}\bm{q})}{\bm{p}^2} n\left(\bm{r},\bm{q} +
\frac{\bm{p}}{2}\right)e^{i\bm{p}(\bm{r} - \bm{x}')}\,.\nonumber
\end{eqnarray}
\noindent Shifting here $\bm{q}\to \bm{q} - \bm{p}/2,$ and
dropping the purely imaginary term proportional to
$(\bm{E}\bm{p}),$ the triple integral becomes purely real, so the
symbol ``${\rm Re}$'' can be omitted. Integrating then over
$\bm{p}$ with the help of the formula
$$\int\frac{d^{\EuScript{D}}\bm{p}}{(2\pi)^{\EuScript{D}}}
e^{i(\bm{p}\bm{x})}\frac{4\pi}{\bm{p}^2} =
\frac{\Gamma(\EuScript{D}/2 -1)}{|\bm{x}|^{\EuScript{D} -
2}\pi^{\EuScript{D}/2 - 1}}\,, \quad \EuScript{D} \ne 2$$ we
obtain
\begin{eqnarray}
C(\bm{x},\bm{x}',t',\omega) =
\frac{2i(2\pi)^{\delta}}{\EuScript{D}(\EuScript{D} - 2)}\frac{e^3
TL^2}{m\omega|\omega|^{\delta}}\iint \frac{d^3 \bm{q}}{(2\pi)^3}
d^3\bm{r}(\bm{E}\bm{q})\frac{n(\bm{r},\bm{q})}{|\bm{r}-\bm{x}'|^{\EuScript{D}
- 2}}\,.\nonumber
\end{eqnarray}
\noindent Finally, substitution of this expression into
Eq.~(\ref{corrfup}) gives low-frequency asymptotic of the power
spectrum of voltage fluctuations
\begin{eqnarray}\label{main}
C_U(\bm{x},\bm{x}',t',\omega) = -
\frac{2i(2\pi)^{\delta}}{\EuScript{D}(\EuScript{D} - 2)}\frac{e^3
TL^2}{\omega|\omega|^{\delta}\Omega} \int d^3\bm{r}
(\bm{E},\overline{\bm{v}}(\bm{r}))\left(
\frac{1}{|\bm{r}-\bm{x}|^{\EuScript{D} - 2}} +
\frac{1}{|\bm{r}-\bm{x}'|^{\EuScript{D} - 2}}\right)\,,
\end{eqnarray}
\noindent where $$\overline{\bm{v}}(\bm{r}) = \Omega\int\frac{d^3
\bm{q}}{(2\pi)^3} \frac{\bm{q}}{m}n(\bm{r},\bm{q})$$ is the local
drift velocity of charge carriers, $\Omega$ denoting the sample
volume. For a crystal in a homogeneous external field,
$\overline{\bm{v}}$ is a function of the crystalline direction,
$$\overline{v}_i = \mu_{ik}E_k\,, \quad i,k=1,2,3,$$ where $\mu_{ik}$
is the charge carrier mobility tensor. To write down the final
expression for the voltage power spectrum, we restore the ordinary
units. Then Eq.~(\ref{main}) takes the form, for $\omega = 2\pi f
>0,$
\begin{eqnarray}\label{mainhom}
C_U(\bm{x},\bm{x}',t',\omega) = - i\,\frac{\varkappa U^2_0}{f^{1 +
\delta}}\,, \quad \varkappa \equiv \frac{e^3c^{\delta}}{\pi\hbar^2
c^3}\mu T g\,,
\end{eqnarray}
\noindent where $$\mu=\mu_{ik}n_i n_k\,, \quad \bm{n} =
\frac{\bm{E}}{|\bm{E}|}\,,$$ $U_0 = |\bm{E}|L$ is the voltage bias
applied to the sample (it is assumed that $\bm{E}
\parallel (\bm{x} - \bm{x}'),$ as is usually the case in practice), and $g$
a geometrical factor
\begin{eqnarray}\label{gfactor}
g = \frac{1}{\EuScript{D}(\EuScript{D} -
2)a^{2\delta}\Omega}\int\limits_{\Omega} d^3\bm{r}\left(
\frac{1}{|\bm{r}-\bm{x}|^{1 - \delta}} +
\frac{1}{|\bm{r}-\bm{x}'|^{1 - \delta}}\right)\,.
\end{eqnarray}
\noindent For $\omega <0,$ the sign of the right hand side in
Eq.~(\ref{mainhom}) is opposite. If Fourier transformation is
defined in a purely real form, i.e., as a decomposition in
$\cos(\omega \tau),$ $\sin(\omega\tau),$ rather than in
$e^{i\omega\tau},$ then the spectral density is also real:
\begin{eqnarray}
C_U(\bm{x},\bm{x}',t',\omega) = \frac{\varkappa U^2_0}{f^{1 +
\delta}}\,.
\end{eqnarray}
\noindent It is seen from Eq.~(\ref{gfactor}) that the $g$-factor
has a pole at $\EuScript{D} = 2,$ which means that the noise
amplitude is unbounded in the limit $\EuScript{D} \to 2.$ It was
already mentioned in Sec.~\ref{calcul} that the case $\EuScript{D}
= 2$ would correspond to the practically unreachable film
thickness $a = d.$ We now see that this case actually cannot be
realized even theoretically. It should be emphasized in this
connection that since in practice the film thickness always
largely exceeds the lattice spacing, only a relatively small
fraction of charge carriers in the film is correlated by the
exchange interaction. Therefore, by continuity, the effective film
dimensionality must be close to $3.$ In other words, the range of
applicability of the developed theory is in any case limited to
flicker noise spectra characterized by sufficiently small values
of $\delta = \gamma-1.$

We mention for future reference that if the sample is an elongated
(say, in $x$-direction) parallelepiped with the leads attached to
its ends, then the $g$-factor can be evaluated approximately as
\begin{eqnarray}\label{gfactorapprox1}
g \approx \frac{2}{\EuScript{D}(\EuScript{D} - 2)a^{2\delta}L w
a}\int\limits_{0}^{L} \frac{w a d x}{x^{1 - \delta}} =
\frac{2}{\delta \EuScript{D}(\EuScript{D} -
2)a^{2\delta}L^{1-\delta}}\,,
\end{eqnarray}
\noindent where $w$ is the sample width, and it is assumed that $a
< w\ll L.$ This is for $\delta >0.$ For $\delta$ close to zero,
this formula is not applicable, because the $x$-integral diverges
near the sample ends. Cutting off the integral at $x\approx w,$
one finds with logarithmic accuracy
\begin{eqnarray}\label{gfactorapprox2}
g \approx \frac{2}{3L w a}\int\limits_{w}^{L} \frac{w a d x}{x} =
\frac{2}{3L}\ln\frac{L}{w}\,, \quad \delta \approx 0\,.
\end{eqnarray}
\noindent We note also that in the $CGS$ system of units, the
$\varkappa$-factor reads
\begin{eqnarray}\label{kappaapprox}
\varkappa \approx 1.62 \cdot 10^{10.48\,\delta - 22}g\mu T\,,
\end{eqnarray}
\noindent where the absolute temperature $T$ is to be expressed in
$^{\circ}{\rm K}.$

\section{Applications}\label{applications}

\subsection{Unboundedness of flicker noise spectrum}\label{unbound}

In this section, a special feature of the derived expression for
the power spectrum, namely, its oddness in frequency, will be
discussed in connection with the problem of observed absence of
frequency limits of the $1/f^{\gamma}$-law. As was mentioned in
Introduction, flicker noise has been detected in a very wide
frequency band  $\sim 10^{-6}\,{\rm Hz}$ to $10^6\,{\rm Hz}.$ This
fact represents one of the essential difficulties for theoretical
explanation, because all physical mechanisms underlying existing
models of flicker noise work in much narrower subbands, and none
of the models suggested so far has been able to explain the
observed plenum of the $1/f$-spectrum.

On the other hand, existence of bounds on this spectrum is
generally believed to be necessary in order to guarantee
finiteness of the total noise power. There is a well-known
argument \cite{flinn} according to which these limits are actually
unnecessary when the flicker noise exponent $\gamma$ is strictly
equal to unity, because the logarithmic divergence of the total
power is not a problem in this case in view of the existence of
natural frequency cutoffs such as the inverse Planck time and
lifetime of Universe. However, this reasoning does not work for
$\gamma \ne 1,$ in which case divergence is a power of the cutoff.
At the same time, the results obtained above reconcile
unboundedness of $1/f$-spectrum with the requirements of
stationarity and finiteness of the total noise power in a quite
natural way. Indeed, using Eq.~(\ref{mainhom}) we find that for
$\delta <1$ the integral
$$\int\limits_{-\infty}^{+\infty}\frac{d\omega}{2\pi} C_U(\bm{x},\bm{x}',t',\omega)
e^{i\omega\tau} = 2i\int\limits_{0}^{+\infty}\frac{d\omega}{2\pi}
C_U(\bm{x},\bm{x}',t',\omega) \sin(\omega\tau) = 2\varkappa U^2_0
\int\limits_{0}^{+\infty}df \frac{\sin(2\pi f \tau)}{f^{1 +
\delta}}$$ converges in both limits $f \to 0$ and $f \to \infty.$
In particular, the singular contribution to the voltage variance
(i.e. to the quantity $C_U|_{\tau = 0}$) vanishes. It is worth to
note also that the total noise power would diverge for $\delta =
1$ even if $\varkappa$ were bounded at $\EuScript{D} = 2.$

Since appearance of odd contributions to the power spectrum is
somewhat unusual in macroscopic fluctuation theory, let us discuss
it in more detail. Under stationary external conditions, the
voltage noise power spectrum (to be denoted below simply as
$C_U(t,t'),$ with the spatial arguments suppressed, for brevity)
must be independent of $t'.$ This is an expression of the noise
stationarity, or, using a term more suitable for the subsequent
discussion, time homogeneity with respect to the macroscopic
system. It is usually realized as the requirement that $C_U(t,t')$
be a function of the difference $t-t' \equiv \tau.$ Since
$C_U(t,t')$ is also symmetric with respect to the interchange $t
\leftrightarrow t',$ an immediate consequence of this is that it
is actually a function of $|\tau|,$ and hence the spectral density
is a real even function of frequency. It is important, on the
other hand, that time homogeneity is not necessarily exhibited by
individual contributions to the total voltage fluctuation,
whatever mechanism of flicker noise generation be. In particular,
this property evidently does not take place at the microscopic
level, i.e., with respect to elementary processes such as charge
carrier trapping, surface or grain boundary scattering, etc.
Stationarity of the macroscopic process emerges usually upon
summation over a large number of individual contributions, so that
this microscopic inhomogeneity turns out to be inconsequential.
However, this summation is not the only way to obtain a stationary
correlation function symmetric in $t,t'.$ Another possibility,
which is realized in the present paper, is that flicker noise may
be a one-particle phenomenon, in the sense that the entire effect
can be ascribed to elementary fluctuations produced by single
charge carriers. In this case the function $C_U(t,t')$ does not
have to depend solely on $|\tau|,$ and as the explicit
calculations of Sec.~\ref{calcul} show, it actually does not.  As
was mentioned above, elementary processes are inhomogeneous in
time, and hence the symmetry with respect to $t \leftrightarrow
t'$ imposes no restriction on the $\tau$-dependence of the
correlation function. The only remaining requirement, namely
reality of the correlation function, implies that contributions to
the spectral density must be real even, {\it or} imaginary odd
functions of frequency [Cf.~Eq.~(\ref{corrg})]. These two cases
correspond to the Fourier decomposition of the function
$C_U(t'+\tau,t')$ in $\cos(\omega\tau)$ and $\sin(\omega\tau),$
respectively, and describe the parts symmetric and antisymmetric
with respect to the difference of its time arguments. Finally,
transition to the statistical distribution removes the
$t'$-dependence of the power spectrum [Cf. discussion after
Eq.~(\ref{diagenk1e2})]. This restores macroscopic time
homogeneity of the correlation function, but leaves the
possibility of being odd with respect to the difference of its
time arguments. In other words, dependence of the power spectrum
on $t'$ shows itself only at microscopic scales, while
macroscopically fluctuations look as if they were homogeneous in
time.

The $1/f$-spectrum derived in the previous section has no lower
frequency cutoff. As to the upper bound, it is given by the
condition $f\ll T$ [see Sec.~\ref{heatbath}], or in the ordinary
units, $f\ll k T/\hbar \approx 10^{11}T\,{\rm Hz},$ with $T$
expressed in $^{\circ}{\rm K}.$ We see that from the practical
point of view, the obtained spectrum has no upper cutoff either.

\subsection{Comparison with experimental data}
\label{comparison}

Let us continue verification of the obtained result and show that
Eq.~(\ref{mainhom}) is in agreement with the other experimentally
established properties of flicker noise. We will first discuss
some general qualitative properties of flicker power spectra,
predicted by Eq.~(\ref{mainhom}), and then give a detailed
quantitative comparison of these predictions with experimental
data.

\subsubsection{Qualitative comparison with the experiment}\label{qualcomparison}

First of all, the power spectrum of quantum electromagnetic
fluctuations, given by Eq.~(\ref{mainhom}), is quadratic in the
applied bias. This is perhaps the most solidly established
property of flicker noise. Second, the noise level is generally
inversely proportional to the sample size. Namely, the $g$-factor
describing dependence of the noise intensity on the sample
dimensions increases roughly as a power of decreasing sample
length or thickness, the exponent depending on the sample geometry
as well as on the effective sample dimensionality $\EuScript{D}$
(which itself depends on the sample thickness.) As to the
dependence of flicker noise amplitude on sample dimensions,
agreement in the literature is not that good. Experiments are
usually arranged so as to prove one of the two main competing
points of view on the flicker noise origin, namely wether it is a
bulk or surface effect. Although this issue is far from being
resolved, there is no doubt that the noise level increases with
decreasing sample size. To be more specific, we note that if
$\delta = {\rm const},$ then Eqs.~(\ref{mainhom}),
(\ref{gfactorapprox1}) tell us that the noise produced by an
elongated sample is proportional to $a^{-2\delta}L^{\delta -1}.$
For experimental verification of this prediction we refer to
\cite{wong1,wong2} which report the results of flicker noise
measurements in various copper films with $a=400\,$\AA $\div
2000\,$\AA, and $L=800\,\mu{\rm m} \div 2000\,\mu{\rm m}.$
According to Ref.~\cite{wong1}, the frequency exponent for samples
on a silicon substrate exceeds noticeably that for equally sized
samples on a sapphire substrate (the ratio of $\delta$'s for the
two cases is about $1.5$). It follows then from the above formulas
that the slope of the noise amplitude considered as a function of
the film thickness is larger for samples on the silicon substrate.
This is indeed the case as is clearly seen from Fig.~3 of
Ref.~\cite{wong1}. Furthermore, since $\delta$ increases for
decreasing $a,$ the slope of the noise amplitude considered as a
function of the sample length is expected to be larger for thicker
samples, but with a less noticeable difference in the slopes since
the length exponent $(\delta-1)$ is less sensitive to variations
in $\delta$ than the thickness exponent $(-2\delta)$ ($\delta$ is
about some tenths in both cases). This is again in agreement with
the observations as is evident from Fig.~4 of Ref.~\cite{wong1}
where the amplitude curve for a $800\,$\AA-thick film is somewhat
steeper than that for a $400\,$\AA-thick film (both on a silicon
substrate).

Next, it is generally agreed that, with other things being equal,
flicker noise is more intensive in semiconductors than in metals,
and this is again in conformity with Eq.~(\ref{mainhom}), because
charge carrier mobility is higher in semiconductors than in
metals, usually by several orders. Unfortunately, determination of
mobility in semiconductors (or semimetals) is a difficult problem,
both theoretically and experimentally, and different experiments
often give significantly different results. By this reason, the
subsequent consideration will be carried out for metals only. Even
in this case careful estimation of the noise level takes some
effort. This is because electron mobilities in thin metal films
commonly used in flicker noise measurements differ essentially
from the corresponding bulk values, varying non-monotonically with
the film thickness, and exhibiting complicated temperature
dependence. Thus, the thicker the film, the more reliable
comparison of theoretical and experimental results. Fortunately,
the modern instrumentation allows measurements in sufficiently
thick samples, electrical transport in which has bulk properties
(usually, effects related to film thickness become important for
$h$ less than a few hundred nanometers). As is well known,
temperature dependence of the electron mobility in this case is
well approximated by the $1/T$ law. Theoretically, this
approximation is valid for $T$ higher than the Debay
characteristic temperature, but in most cases it is practically
applicable already for $T\gtrsim 50^{\circ}{\rm K}.$ Furthermore,
the effective dimensionality $\EuScript{D}\to 3$ in thick samples.
Therefore, it follows from Eq.~(\ref{mainhom}) that in
sufficiently thick samples the frequency exponent $\gamma = 1 +
\delta \approx 1,$ and the flicker noise level is temperature
independent. This conclusion is confirmed, e.g., by the results of
Ref.~\cite{massiha} where $1/f$ noise was measured in $2.44\,{\rm
\mu m}$ thick metal films, which is quite sufficient for bulk
treatment of the sample conduction. According to Fig.~5 of
Ref.~\cite{massiha}, the flicker noise level is constant for
$T\gtrsim 50^{\circ}{\rm K}$ indeed, and $\gamma$ is found to be
about $1.01$ for most samples.\footnote{As mentioned in
Ref.~\cite{massiha}, at very high current densities exceeding
$1.1\cdot 10^6\,{\rm A}/{\rm cm^2},$ the samples undergone
structural defects resulting in somewhat higher values of
$\gamma.$ Unfortunately, the authors of \cite{massiha} did not
specify the metals used in their experiments, which makes
quantitative comparison with Eq.~(\ref{mainhom}) impossible.} In
the opposite case of extremely thin films, conductivity is
approximately independent of temperature. For instance,
resistivity of a $5\,{\rm nm}$ thick gold film varies from the
value $2.85\cdot 10^{-6}\Omega\cdot {\rm m}$ at
$T=50\,^{\circ}{\rm K}$ to $3.0\cdot 10^{-6}\Omega\cdot {\rm m}$
at $T=275\,^{\circ}{\rm K},$ i.e., only by about $5\%$ \cite{pov}.
If $\delta = 3 - \EuScript{D}$ were constant, this would mean that
the noise magnitude is a linear function of temperature in this
case. However, $\delta$ is itself temperature dependent in thin
films, namely, it is expected to decrease with increasing
temperature, because thermal effects destroy the exchange
correlations, thus raising the effective dimensionality of the
sample.

Finally, it was found in \cite{wong1} that the frequency exponent
in some cases depends also on the sample length, though much more
weakly than on the sample thickness. This dependence cannot be
explained from the point of view of the developed theory, so its
appearance can serve as an indication on the limits of
applicability of the theory.

\subsubsection{Quantitative comparison with the
experiment}\label{quantcomparison}

In order to compare the absolute value of the noise spectral
density given by Eq.~(\ref{mainhom}) with experimental data, we
refer to the results of flicker noise measurements performed by
Wong, Cheng and Ruan \cite{wong1}, which were used already in the
above qualitative analysis, and by Voss and Clarke \cite{voss1}.
Before going into detailed comparison, let us mention the
following important circumstance. As is seen from
Eq.~(\ref{mainhom}), the noise amplitude is very sensitive to the
value of $\delta.$ This parameter appears, in particular, in the
exponent of the ratio $(c/f a^2)$ which is normally very large.
Indeed, for $f = 1\,{\rm Hz}$ and $a = 10^{-5}\,{\rm cm},$ this
ratio is equal to $3 \cdot 10^{20}.$ Therefore, an error as small
as $0.01$ in the value of $\delta$ results in the extra factor of
$10^{0.2}\approx 1.6$ in the amplitude. At the same time, $\delta$
is usually measured with the accuracy of a few hundredths at best,
so the calculation carried out below is actually an
order-of-magnitude estimation of the noise level.

\paragraph{Wong, Cheng and Ruan.} In this work, flicker noise power
spectra were measured in the case of copper films of various
thickness and length, sputtered on sapphire and oxidized silicon
substrates. To illustrate the scheme of the calculation, let us
take as an example the case of film with $L = 1200\,\mu{\rm m},$
$a = 1,2\cdot 10^{-5}\,{\rm cm},$ deposited on the sapphire wafer.
According to Fig.~1 of \cite{wong1}, samples of this thickness
have conductivity $\sigma = 5.9 \cdot 10^6\,\Omega^{-1}{\rm
m^{-1}}.$ Using the relation $\mu = \sigma/en$ where $n = 8\cdot
10^{22}\,{\rm cm}^{-3}$ is the charge carrier concentration in
copper, one finds the electron mobility $\mu = 1.4\cdot 10^3$
units CGS. Next, according to Fig.~5 of \cite{wong1}, the
frequency exponent $\gamma$ in the case under consideration is
equal to $1.1,$ hence, $\delta = \gamma - 1 = 0.1.$ Substituting
these values into the formula (\ref{gfactorapprox1}) gives $g =
510$ units CGS. Finally, putting this and\footnote{This value of
$T$ as well as the accuracy of $\delta,$ mentioned below, are
communicated to the author by Prof. H.~Wong.} $T =
300\,^{\circ}{\rm K}$ in Eq.~(\ref{kappaapprox}), one obtains
$\varkappa = 3.8\cdot 10^{-13}.$ This value is to be compared with
the measured value $\varkappa_{\rm exp} = 6\cdot 10^{-13}$ given
in Fig.~3 of Ref.~\cite{wong1} (where it is called normalized
noise amplitude). The experimental error of $\delta$ is about
$0.02,$ which implies an ambiguity by the factor of $2.5$ in the
calculated value of $\varkappa.$

It was found in \cite{wong1} that sufficiently thin copper films
are characterized by a pronounced dependence of the frequency
exponent on the sample length. Figure~5 of \cite{wong1} shows that
the thinner the film, the stronger this dependence: for $L$
increasing from $800\,\mu{\rm m}$ to $2000\,\mu{\rm m},$ $\gamma$
decreases by only $0.05$ in the case of $a = 800\,$\AA, and by
$0.15$ in the case of $a = 400\,$\AA. As was already mentioned
above, the present theory is unable to explain this dependence,
and hence the corresponding experimental data are beyond the scope
of applicability of the theory. By this reason, the quantitative
comparison below is carried out only for films with $a \geqslant
800\,$\AA. The calculated and measured values of $\varkappa$ for
films of various thickness on sapphire and silicon substrates are
collected in Table~I and Table~II, respectively, together with the
other parameters involved in the calculation. The quantities $\mu,
g ,\varkappa$ are given in the CGS system of units.

It is seen from these tables that the calculated and measured
values of $\varkappa$ for films on sapphire wafer agree within the
experimental error for all $a.$ For films on silicon wafer, the
agreement within the experimental error is found for $a = 1200\,$
\AA and $a = 1600\,$\AA.  In the case of $a = 800\,$\AA, where
dependence of $\gamma$ on $L$ is still noticeable, the theory
somewhat overestimates the noise level.

\paragraph{Voss and Clarke.}
In the work \cite{voss1}, flicker noise was measured in thin metal
films evaporated or sputtered on glass substrates. The information
provided in this paper is sufficient for estimation of the noise
intensity in the gold film shown in Fig.~2 of \cite{voss1}. This
was an elongated sample with $h = 25\,{\rm nm},$ $w = 8\,{\rm\mu
m},$ $l = 625\,{\rm\mu m},$ biased at $U_0 = 0.81\,{\rm V},$ and
operated at about $40^{\circ}{\rm K}$ above room temperature.
Unfortunately, the exact value of $\gamma$ is not given for this
case by the authors, who mentioned only that it is close to $1.$
In view of what have been said about dependence of the noise
amplitude on $\delta,$ this implies a large amount of uncertainty
in the theoretical estimation.\footnote{The measured spectrum
presented in \cite{voss1} does not actually fit the
$1/f^{\gamma}$-law even at low frequencies, because of the
low-frequency roll off of the amplifier and capacitor used to
improve impedance match for low-resistance samples. As a result,
after {\it subtraction} of the background noise contribution, the
low-frequency part of the corrected spectrum curve in Fig.~2 of
Ref.~\cite{voss1} goes {\it above} the measured values.} Yet,
assuming that $\delta \approx 0$ indeed, we substitute the sample
dimensions in Eq.~(\ref{gfactorapprox2}) and find $g = 46\,{\rm
cm}^{-1}.$ Next, in order to determine conductivity, we use the
$I-V$ characteristic of the given gold sample, shown in Fig.~3 of
\cite{voss1}. According to this figure, the sample resistance was
about $100\, \Omega.$ Taking into account the sample dimensions
given above, this implies that $\sigma =1.2\cdot 10^{6}\,{\rm
\Omega^{-1} m^{-1}}.$ It should be mentioned that this value is
approximately six times lower than that obtained in more recent
studies of electrical transport in thin films. For instance,
according to Ref.~\cite{bieri} conductivity of a $25\,{\rm nm}$
thick, $15\,{\rm \mu m}$ wide gold film obtained by a
laser-improved deposition of nanoparticle suspension, is $7.1
\cdot 10^{6}\,{\rm \Omega^{-1} m^{-1}}.$ The same value can be
obtained also indirectly using the data given in
Refs.~\cite{chen,pov}. According to \cite{chen}, the conductivity
of gold is $75\%$ to $85\%$ of its bulk value for $h = 100\,{\rm
nm},$ depending on the choice of the substrate, and decreases
below that value approximately linearly with decreasing thickness.
On the other hand, according to Ref.~\cite{pov} conductivity drops
to about $3\cdot 10^{5}\,{\rm \Omega^{-1} m^{-1}}$ for $h =
5\,{\rm nm}.$ One readily finds from this that for $h = 25\,{\rm
nm},$ $\sigma = (6.5 \div 7.5) \cdot 10^{6}\,{\rm \Omega^{-1}
m^{-1}}.$ Presumably, this difference in the values of
conductivity is to be attributed to the quality of film
deposition. Substitution of $\sigma =1.2\cdot 10^{6}\,{\rm
\Omega^{-1} m^{-1}}$ and $n = 5.9\cdot 10^{22}\,{\rm cm}^{-3}$ in
the relation $\mu=\sigma/en$ yields the electron mobility $\mu =
390$ units CGS. Then Eq.~(\ref{kappaapprox}) gives $\varkappa =
9.6\cdot 10^{-16}$ (for $T = 330\,^{\circ}{\rm K}$). Putting this
together with the bias value given above in Eq.~(\ref{mainhom}),
we find $|C_U| = 6.3\cdot 10^{-16}\,{\rm V^2/Hz}$ for the
frequency $f = 1\,{\rm Hz},$ which is to be compared with the
experimental value $C_U \approx 10^{-15}\,{\rm V^2/Hz}\,.$

\section{Discussion and Conclusions}

We have shown that, staying within the one-particle picture of
flicker noise generation by quantum electromagnetic fluctuations,
backreaction of the conducting medium on the fluctuating field of
the charge carrier can be described phenomenologically as an
effective reduction of the system dimensionality. Using the
dimensional continuation technique, we have found that the
backreaction affects both the frequency dependence and the
magnitude of the noise spectrum. Namely, the frequency exponent in
the $1/f^{\gamma}$-asymptotic of the fluctuation power spectrum in
a conducting sample is found to be $1 + \delta,$ where $\delta = 3
- \EuScript{D},$ $\EuScript{D}$ being the effective dimensionality
of momentum space, while dependence of the noise amplitude on
$\delta$ is given by Eqs.~(\ref{mainhom}), (\ref{gfactor}).
Although introduced initially as a momentum space characteristic,
$\delta$ thus relates the noise amplitude to geometric properties
of the sample in the ordinary $3$-dimensional coordinate space
[Cf. Eqs.~(\ref{gfactor}), (\ref{gfactorapprox1})]. It was
demonstrated in Sec.~\ref{qualcomparison} that the experimentally
observed dependence of the noise amplitude on the sample geometry
for a given $\delta$ is adequately described by
Eq.~(\ref{gfactorapprox1}). The way $\delta$ itself depends on the
sample thickness and system temperature, as expected from its
definition, is also confirmed by observations. However, the value
of $\delta$ cannot be predicted within the developed approach. In
other words, $\delta$ plays the role of a phenomenological
parameter of the theory. The noise amplitude turns out to be
notably sensitive to the value of $\delta$: We saw in
Sec.~\ref{quantcomparison} that for a $10^{-5}\,{\rm cm}$-thick
film, an increase of $0.1$ in $\delta$ raises the noise level by
about two orders. This perfectly agrees with the observed rise of
the noise level in samples with $\gamma
> 1$ in comparison\footnote{It is meant that compared are the
noise levels at a fixed frequency, rather than the spectra
themselves (in the latter case, of course, comparison would be
meaningless).} with the predictions of the Hooge's empirical
formula \cite{hooge1} obtained for $\gamma=1.$ In fact, the
quantitative comparison carried out in Sec.~\ref{quantcomparison}
shows that the calculated and measured values of the parameter
$\varkappa$ coincide within the experimental error. Deviations
between the theory and experiment become noticeable only in films
with a pronounced dependence of the frequency exponent on the
sample length. It was mentioned in Sec.~\ref{qualcomparison} that
this dependence cannot be explained from the point of view of the
developed theory, so its appearance indicates the limits of
applicability of the theory. Although this issue is beyond the
scope of the present approach, it is yet worth to comment on the
possible origin of these deviations. According to
Ref.~\cite{wong1}, dependence of the frequency exponent on the
sample length is noticeable in sufficiently thin films. At the
same time, the fact that films deposited on different substrates
exhibit different noise characteristics clearly shows that the
boundary conditions affect significantly the mechanism of noise
generation. From the theoretical point of view, this is reflected
in the essential role played by the assumption that the film is
plane-parallel in the derivation of the photon propagator [Cf.
discussion below Eq.~(\ref{planepar})]. On the other hand, this
assumption is violated to some extent by the surface roughness of
the film. In fact, the authors of \cite{wong1} emphasize that the
electric properties of the films used in their experiments are
affected by the surface roughness, especially in the case of thin
films. Thus, the surface roughness is a possible reason for the
above-mentioned deviations in the values of $\varkappa.$

Finally, as was shown in Sec.~\ref{unbound}, the obtained results
explain the observed unboundedness of the flicker noise spectra,
resolving naturally the problem of divergence of the total noise
power. Together with the demonstrated qualitative and quantitative
agreement of the results with experimental data, this suggests
that quantum electromagnetic fluctuations is the source of flicker
noise in metal films.

\begin{appendix}

\section{Gauge independence of the power spectrum}

Consider the theory of interacting scalar and electromagnetic
fields described by the action
$$S = S_{\phi} + S_{A}\,,$$ where $S_{\phi}$ is given by
Eq.~(\ref{action}), and
$$S_{A} = - \frac{1}{4}{\displaystyle\int} d^4 x F_{\mu\nu} F^{\mu\nu}
+ S_{gf}\,, \quad S_{gf} = \frac{1}{2\alpha}{\displaystyle\int}
d^4 x~(\partial_{\mu}A^{\mu})^2\,,\quad F_{\mu\nu} =
\partial_{\mu} A_{\nu} - \partial_{\nu} A_{\mu}\,.$$ For arbitrary constant parameter
$\alpha,$ the gauge fixing term describes the generalized Lorentz
gauge. Let us introduce the generating functional of Green
functions
\begin{eqnarray}\label{gener1}
Z[J,\eta,\eta^*] = \int dA d\phi d\phi^* \exp\left\{i\left(S +
\int d^4x[J^{\mu}A_{\mu} + \eta^*\phi + \eta\phi^*]
\right)\right\}\,,
\end{eqnarray}
\noindent where $J,\eta,\eta^*$ denote sources for the fields
$A,\phi^*,\phi,$ respectively. Vanishing of $Z$ under the gauge
variation of the functional integral variables
\begin{eqnarray}\label{gaugetransform}
\delta A_{\mu} =
\partial_{\mu}\xi(x)\,, \quad \delta\phi = ie\xi(x)\phi\,,
\quad \delta\phi^* = -ie\xi(x)\phi^*\,,
\end{eqnarray}
\noindent with $\xi(x)$ a small gauge function, leads to the Ward
identity
\begin{eqnarray}\label{ward}
- i\partial_{\mu} J^{\mu}(y)Z +
\frac{\Box}{\alpha}\partial_{\mu}\frac{\delta Z}{\delta
J_{\mu}(y)} + ie\eta^*(y)\frac{\delta Z}{\delta \eta^*(y)} -
ie\eta(y)\frac{\delta Z}{\delta \eta(y)} = 0\,.
\end{eqnarray}
\noindent Since we are interested in the connected contribution to
the correlation function, we rewrite this identity for the
generating functional of connected Green functions, $W = - i\ln
Z,$
\begin{eqnarray}\label{wardc}
- \partial_{\mu} J^{\mu}(y) +
\frac{\Box}{\alpha}\partial_{\mu}\frac{\delta W}{\delta
J_{\mu}(y)} + ie\eta^*(y)\frac{\delta W}{\delta \eta^*(y)} -
ie\eta(y)\frac{\delta W}{\delta \eta(y)} = 0\,.
\end{eqnarray}
\noindent The consequence of this equation we need is obtained by
functional differentiation with respect to $\eta,\eta^*,$ and
twice with respect to $J,$ with all the sources set equal to zero
afterwards,
\begin{eqnarray}&&
\frac{\Box^y}{\alpha}\partial^y_{\mu}\frac{\delta^5 W}{\delta
J_{\mu}(y)\delta J_{\alpha}(x)\delta
J_{\beta}(x')\delta\eta(z)\delta\eta^*(z')} + ie\delta^{(4)}(y -
z')\frac{\delta^4 W}{\delta J_{\alpha}(x)\delta
J_{\beta}(x')\delta\eta(z)\delta \eta^*(y)} \nonumber\\&& -
ie\delta^{(4)}(y - z)\frac{\delta^4 W}{\delta J_{\alpha}(x)\delta
J_{\beta}(x')\delta \eta(y)\delta\eta^*(z')} = 0\,. \nonumber
\end{eqnarray}
\noindent Fourier transform of this identity with respect to $y$
reads
\begin{eqnarray}\label{wardc1}&&
\frac{k'^2}{\alpha}k'_{\mu}\int d^4 y e^{-ik' y}\frac{\delta^5
W}{\delta J_{\mu}(y)\delta J_{\alpha}(x)\delta
J_{\beta}(x')\delta\eta(z)\delta\eta^*(z')} \nonumber\\ &&=
e(e^{-ik' z'} - e^{-ik' z}) \frac{\delta^4 W}{\delta
J_{\alpha}(x)\delta J_{\beta}(x')\delta\eta(z)\delta
\eta^*(z')}\,.
\end{eqnarray}
\noindent The argument of the Fourier transform is purposely
denoted here by $k'$ to stress that the left hand side of this
equation corresponds to the variation of the Green function we
dealt with in Sec.~\ref{calcul}, under gauge variation of the
external field. Indeed, the longitudinal part of the photon
propagator in the generalized Lorentz gauge has the form
\begin{eqnarray}\label{long}
D^l_{\mu\nu}(k) = - \alpha\frac{k_{\mu}k_{\nu}}{k^4}\,.
\end{eqnarray}
\noindent Therefore, contraction with the factor $k'^2
k'_{\mu}/\alpha$ is equivalent to amputation of the photon
propagator attached to the $y$ vertex, followed by contraction of
this vertex with $k'_{\mu}.$ Exactly the same result is obtained
under the gauge variation of the external field coming into this
vertex. The only difference with the Green function we considered
in Sec.~\ref{calcul} is that the external scalar lines in
Eq.~(\ref{wardc1}) are the particle propagators. To promote them
into particle amplitudes, according to the standard rules,
Eq.~(\ref{wardc1}) is to be Fourier transformed with respect to
the variables $z,z',$ and then multiplied by
$\psi(\bm{q})\psi^*(\bm{q}')(m^2 - q^2)(m^2 - q'^2),$ where the
arguments $q,q'$ of the Fourier transformations with respect to
$z,z'$ are to be taken eventually on the mass shell. But these
operations give zero identically when applied to the right hand
side of Eq.~(\ref{wardc1}), because each of the factors $e^{-ik'
z'},$ $e^{-ik' z}$ makes the corresponding particle propagator
nonsingular on the mass shell. For instance, the first term in
Eq.~(\ref{wardc1}) gives rise to the contribution of the form
$(m^2 - q'^2)D^{\phi}(q' + k')$ times terms nonsingular on the
mass shell. For $k' \ne 0,$ the function $D^{\phi}(q' + k')$ is
also nonsingular at $q'^2=m^2,$ and hence this contribution
vanishes on the mass shell.

Thus, the correlation function is invariant under gauge
transformations of the external field, which are part of the gauge
freedom in the theory. The other part is related to the explicit
dependence of the photon propagator on the choice of the gauge
conditions used to fix the gauge invariance of the action. As is
well known, it is longitudinal part of the propagator that depends
on the gauge. Let us first consider the simples case of
Lorentz-invariant gauges. Then the most general form of the
longitudinal part is given by Eq.~(\ref{long}) in which $\alpha$
is to be regarded as an arbitrary function of $k^2.$ It is not
difficult to see that variations of $\alpha(k^2)$ do not affect
the observable quantities. Recall, first of all, that we are
interested ultimately in the fluctuations of gauge-invariant
quantities such as the electric field strength. The
$\alpha$-independence of these quantities is a direct consequence
of their gauge invariance, because variations of $\alpha(k^2)$
give rise to terms that are pure gradients with respect to the
spacetime arguments $x,x',$ as is easily verified by substituting
the expression (\ref{long}) in place of one or two photon
propagators in Eq.~(\ref{diagen}). Then, if the vector potential
contribution to the field strength is negligible, as is the case
in our nonrelativistic calculation (recall the condition
$|\bm{q}|\ll m$ used throughout), the voltage correlation function
can be found by integrating the correlation function for the field
strength with respect to $\bm{x},\bm{x}'$ using the relation
$\bm{E} = - \bm{\nabla}A_0.$

More generally, the longitudinal part of the photon propagator in
a Lorentz non-invariant gauge has the form, in coordinate space,
$$D^l_{\mu\nu}(x) = \partial_{\mu}\chi_{\nu}(x)
+ \partial_{\nu}\chi_{\mu}(x)\,,$$ where $\chi_{\mu}(x)$ is an
arbitrary function of spacetime coordinates. If the Lorentz index
of the spacetime derivative in $\partial_{\mu}\chi_{\nu}(x)$ is
left free after combining the Feynman diagrams in Fig.~\ref{fig2},
then this term leads to a gradient contribution to the two-point
function of electromagnetic field, and, as before, does not
contribute to the voltage correlation function. On the other hand,
if the spacetime derivative is contracted with the interaction
vertex, then the contribution of $\partial_{\mu}\chi_{\nu}(x)$ to
the two-point function is not pure gradient, so the above argument
does not work. Nevertheless, it can be shown that all such terms
cancel each other in the complete expression for the correlation
function. However, this requires examination of the complete set
of Feynman diagrams, which complicates the proof. To avoid this
complication, we will prove gauge independence of the
low-frequency asymptotic of the power spectrum only, which is
quite sufficient for our purposes. To this end, we use the
following identity expressing invariance of the particle action
under the transformation (\ref{gaugetransform})
$$\frac{\delta S_{\phi}}{\delta \phi(x)}ie\phi(x) -
\frac{\delta S_{\phi}}{\delta \phi^*(x)}ie\phi^*(x) -
\partial^{x}_{\mu}\frac{\delta S_{\phi}}{\delta A_{\mu}(x)} = 0\,.$$
Differentiating this identity twice with respect to $\phi(y),$
$\phi^*(z)$ and setting $A_{\mu}=0$ afterwards yields
$$\partial^{x}_{\mu}\frac{\delta^3 S_{\phi}}
{\delta A_{\mu}(x)\delta\phi(y)\delta\phi^*(z)} = \frac{\delta^2
S_{\phi}}{\delta \phi(x)\delta \phi^*(z)}ie\delta^{(4)}(x-y) -
\frac{\delta^2 S_{\phi}}{\delta
\phi^*(x)\delta\phi(y)}ie\delta^{(4)}(x - z)\,.$$ The left hand
side here is just the interaction vertex contracted with the
derivative coming from the term $\partial_{\mu}\chi_{\nu}(x).$ It
follows that the result of this contraction is expressed through
the inverse charged particle propagator. When built into the
Feynman diagrams in Fig.~\ref{fig2}, the latter is integrated with
either the function $\psi(z)$ (the external line), or the particle
propagator. In the first case it gives zero by virtue of the
definition of free particle states, while in the second it cancels
the particle propagator. At the same time, as we saw in
Sec.~\ref{calcul}, it is singularity of the particle propagator
that is responsible for the occurrence of $1/\omega$ contribution
to the function $G(\bm{x},\bm{x}',t',\omega).$ Hence, longitudinal
part of the photon propagator does not contribute to the leading
low-frequency term of the power spectrum of electromagnetic
fluctuations.

Thus, gauge-independence of our results expressed by
Eqs.~(\ref{main}), (\ref{mainhom}) is proved completely.

\section{Photon propagator in thin metal film at $T=0.$}

For the sake of clarity, we keep track of the factor $a$ in the
formulas of the present section. As was mentioned in
Sec.~\ref{exchangeint}, the only component of the photon
propagator, relevant in the calculation of the two-point
correlation function, is $D_{00}(x-x') = i\langle \EuScript{T}
\hat{A}_{0}(x)\hat{A}_{0}(x')\rangle.$ This important fact will be
proved below in the case of the axial gauge (\ref{axial}) which
was seen to be particularly convenient in describing the damping
of the temporal photon propagation in the $3$-direction. Let the
unit vector in this direction be denoted by $n,$ $$n_{\mu} = -
\eta_{3\mu}.$$ To determine the form of the photon propagator at
zero temperature
\begin{eqnarray}\label{zerotprop}
D_{\mu\nu}(x-x')|_{T=0} \equiv D^0_{\mu\nu}(x-x') = \langle
0|\EuScript{T} \hat{A}_{\mu}(x)\hat{A}_{\nu}(x')|0 \rangle\,,
\end{eqnarray}
\noindent we go over to momentum space
\begin{eqnarray}\label{mspaced0}
D^0_{\mu\nu}(x) = \int\frac{d^4 k}{(2\pi)^4}
e^{-ikx}D^0_{\mu\nu}(k)\,,
\end{eqnarray}
\noindent and note that $D^0_{\mu\nu}(k)$ is a symmetric rank-two
tensor that can be built only from the vectors $n_{\mu,}$
$k_{\mu},$ and the Minkowski tensor $\eta_{\mu\nu}.$ Therefore, it
has the following general structure
\begin{eqnarray}
D^0_{\mu\nu}(k) = \left\{\eta_{\mu\nu} + an_{\mu}n_{\nu} + b
(k_{\mu}n_{\nu} + k_{\nu}n_{\mu}) + ck_{\mu}k_{\nu}
\right\}\Delta\,,\nonumber
\end{eqnarray}
\noindent where $a,b,c,$ and $\Delta$ are some scalars built from
$n_{\mu,}$ $k_{\mu},$ i.e., functions of the two invariant
combinations $(nk) = -k_3$ and $k^2.$ According to discussion in
Sec.~\ref{exchangeint}, the electromagnetic field operator
$\hat{A}_{\mu}$ is constrained by $\hat{A}_3 = 0,$
$\partial\hat{A}_0/\partial x_3 = 0,$ which imply the following
conditions on the form of the photon propagator
\begin{eqnarray}\label{conditions}
D_{3\nu}(x) = 0\,, \quad \frac{\partial D_{0\nu}(x)}{\partial x_3}
= 0\,.
\end{eqnarray}
\noindent Written out in components, the first of these conditions
reads
\begin{eqnarray}
(bk_0 + ck_3k_0)\Delta &=& 0\,, \nonumber \\
(bk_i + ck_3k_i)\Delta &=& 0\,, \quad i =1,2\,,\nonumber  \\
(- 1 + a + 2bk_3 + ck^2_3)\Delta &=& 0\,.\nonumber
\end{eqnarray}
\noindent It follows that $$b = - ck_3 = c(nk)\,, \quad c =
\frac{a-1}{k^2_3} = \frac{a-1}{(nk)^2}\,,$$ so
\begin{eqnarray}
D^0_{\mu\nu}(k) = \left\{\eta_{\mu\nu} + an_{\mu}n_{\nu} +
(a-1)\frac{(k_{\mu}n_{\nu} + k_{\nu}n_{\mu})}{(nk)} +
(a-1)\frac{k_{\mu}k_{\nu}}{(nk)^2} \right\}\Delta\,.
\end{eqnarray}
\noindent Then the second of the conditions (\ref{conditions})
gives
\begin{eqnarray}
\left\{(nk)\eta_{0\nu} + (a-1)k_0n_{\nu} +
(a-1)\frac{k_0k_{\nu}}{(nk)} \right\}\Delta = 0\,.\nonumber
\end{eqnarray}
\noindent The $3$-component of this condition is satisfied
identically, while the other three yield
\begin{eqnarray}\label{scond1}
\left\{(nk) + (a-1)\frac{k^2_0}{(nk)} \right\}\Delta &=& 0\,, \\
(a-1)\frac{k_0k_i}{(nk)} \Delta &=& 0\,, \label{scond2}\quad i
=1,2\,.
\end{eqnarray}
\noindent Since the photon propagator does depend on
$x_0,x_1,x_2,$ it follows form Eq.~(\ref{scond2}) that $a = 1,$
and then Eq.~(\ref{scond1}) gives $k_3\Delta = 0$ implying that
the function $D^0_{\mu\nu}(x)$ is independent of $x_3.$ Thus,
\begin{eqnarray}\label{d0gen}
D^0_{\mu\nu}(k) = \left(\eta_{\mu\nu} +
n_{\mu}n_{\nu}\right)\Delta\,.
\end{eqnarray}
\noindent It remains only to find the quantity $\Delta.$ To this
end, we will calculate the $00$-component of the propagator
explicitly.

Under the condition $\partial\hat{A}_0/\partial x_3 = 0,$ the
normal mode decomposition of the scalar potential operator reads
\begin{eqnarray}\label{operator}&&
\hat{A}_0(\bm{x},t) =
N\sum\limits_{k_1,k_2}\frac{1}{\sqrt{2\omega_k}}
\left(\hat{a}_{\bm{k}}\exp\{i(-\omega_k t + k_1 x_1 + k_2 x_2 )\}
+ {\rm H.c.}\right)\,, \quad \omega_k = |\bm{k}|\,, \nonumber
\end{eqnarray}
\noindent where $N$ is a normalization factor to be determined
below, and $\hat{a}_{\bm{k}}$ is the annihilation operator of a
``temporal'' photon with the wave vector $\bm{k} = (k_1,k_2,0).$
As is known from quantum electrodynamics, the commutator of this
operator with its Hermitian conjugate is
$\hat{a}_{\bm{k}}\hat{a}^{\dagger}_{\bm{k}} -
\hat{a}^{\dagger}_{\bm{k}}\hat{a}_{\bm{k}} = -1\,.$ Substituting
the expression for $\hat{A}_0$ in Eq.~(\ref{zerotprop}), one finds
$$D^0_{00}(\bm{x},t)
= - N^2\sum\limits_{\bm{k}}\frac{i}{2|\bm{k}|} \exp\{i(-\omega_k
|t| + \bm{k}\bm{x})\}\,.$$ As always, the factor $N$ depends on
the choice of a ``box'' in which the field is quantized. Suppose
that this box is a rectangular in $(x_1,x_2)$-plane, its sides
being parallel to the coordinate axes. For a sufficiently large
box, summation over $\bm{k}$ can be replaced by integration over
$Sd^2\bm{k}/(2\pi)^2,$ where $S$ is the box area,
$$D^0_{00}(\bm{x},t) = - iN^2 S\int\frac{d^2\bm{k}}{(2\pi)^2}
\frac{e^{i(-\omega_k |t| + \bm{k}\bm{x})}}{2|\bm{k}|} = N^2
S\int\limits_{-\infty}^{+\infty} \frac{dk_0}{2\pi}
\int\frac{d^2\bm{k}}{(2\pi)^2} \frac{e^{i(- k_0 t +
\bm{k}\bm{x})}}{k^2_0 - \bm{k}^2 + i0}\,.$$ $N$ can be found by
calculating the electric potential produced by a resting point
charge, $e.$ If this charge is at the origin of the coordinate
system, then the potential at the point $\bm{x}= (x_1,x_2,0)$ is
given by the well-known formula
$$A_0(\bm{x}) = - e\int\limits_{-\infty}^{+\infty}
dt D^0_{00}(\bm{x},t)\,.$$ Substitution of the explicit expression
for $D^0_{00}(\bm{x},t)$ yields
\begin{eqnarray}\label{planepar}
A_{0}(\bm{x}) = eN^2
S\int\frac{d^2\bm{k}}{(2\pi)^2}\frac{e^{i\bm{k}\bm{x}}}{\bm{k}^2}\,.
\end{eqnarray}
\noindent On the other hand, since this potential is independent
of $x_3,$ the same result will be obtained if the film is replaced
by an infinite set of identical parallel films adjacent to each
other. This is allowed by our assumption that the film is
plane-parallel. In other words, $A_0(\bm{x})$ can be represented
as a superposition of the Coulomb potentials produced by an
infinite sequence of vertically aligned point charges, spaced at
the distance $a,$ or equivalently, by a charge distribution with
density $\rho(x_1,x_2,x_3) = (e/a)\delta^{(2)}(x_1,x_2).$ Fourier
transform of the latter is $\rho(\bm{k}) = (e/a)2\pi\delta (k_3),$
hence
$$A_0(\bm{x}) = \int d^3 \bm{x}'
\frac{\rho(\bm{x}')}{|\bm{x} - \bm{x}'|} = \int\frac{d^3 \bm{k}
}{(2\pi)^3}\rho(\bm{k})\frac{4\pi e^{i\bm{k}\bm{x}}}{\bm{k}^2} =
\frac{4\pi e}{a}\int \frac{d^2\bm{k}}{(2\pi)^2}
\frac{e^{i\bm{k}\bm{x}}}{\bm{k}^2}\,,$$ and comparison with the
preceding formula gives $$N = \sqrt{\frac{4\pi}{Sa}}\,.$$ The
$00$-component of the photon propagator thus takes the form
\begin{eqnarray}
D^0_{00}(x) = \int\frac{d^4 k}{(2\pi)^4} e^{-ikx}\Delta\,, \quad
\Delta = \frac{8\pi^2}{a}\frac{\delta(k_3)}{k^2 + i0}\,. \nonumber
\end{eqnarray}
\noindent  Putting this $\Delta$ in Eqs.~(\ref{d0gen}),
(\ref{mspaced0}), we arrive finally at the following expression
for the zero-temperature photon propagator
\begin{eqnarray}
D^0_{\mu\nu}(x) = \left(\eta_{\mu\nu} +
n_{\mu}n_{\nu}\right)\frac{1}{a}\int\frac{d^3 k}{(2\pi)^3}
e^{-ikx}\frac{4\pi}{k^2 + i0}\,.\nonumber
\end{eqnarray}
\noindent

\end{appendix}

\pagebreak

\begin{table}
\begin{tabular}{cccccc}
\hline
  \hspace{0,3cm} $a,$ \AA \hspace{0,3cm}
  & \hspace{0,5cm} $\delta$ \hspace{0,5cm}
  & \hspace{0,3cm} $\mu\times 10^{-3}$ \hspace{0,3cm}
  & $G\times 10^{-2}$
  & \hspace{0,3cm} $\varkappa_{\rm th}\times 10^{13}$ \hspace{0,3cm}
  & $\varkappa_{\rm exp} \times 10^{13}$   \\
\hline
800 & 0.14 & 1.1 & 9.8 & 15 & 9 \\
1200 & 0.10 & 1.4 & 5.1 & 3.8 & 6 \\
1600 & 0.10 & 2.2 & 4.7 & 5.5 & 5 \\
\hline
\end{tabular}
\caption{Calculated ($\varkappa_{\rm th}$) and measured
($\varkappa_{\rm exp}$) values of $\varkappa$  for copper films of
various thickness on sapphire substrates. $\mu, g ,\varkappa$ are
given in the CGS system of units.} \label{table1}
\end{table}

\begin{table}
\begin{tabular}{cccccc}
\hline
  \hspace{0,3cm} $a,$ \AA \hspace{0,3cm}
  & \hspace{0,5cm} $\delta$ \hspace{0,5cm}
  & \hspace{0,3cm} $\mu\times 10^{-3}$ \hspace{0,3cm}
  & $G\times 10^{-2}$
  & \hspace{0,3cm} $\varkappa_{\rm th}\times 10^{12}$ \hspace{0,3cm}
  & $\varkappa_{\rm exp} \times 10^{12}$   \\
\hline
800 & 0.21 & 0.9 & 32 & 22 & 2 \\
1200 & 0.17 & 1.2 & 14 & 4.9 & 2 \\
1600 & 0.14 & 2.0 & 8 & 2.3 & 1 \\
\hline
\end{tabular}
\caption{Same for copper films on silicon substrates.}
\label{table2}
\end{table}

\pagebreak

. \vspace{10cm}

\pagebreak

\pagebreak

\end{document}